\documentclass[prb,aps,twocolumn,amsmath,amssymb,floatfix,citeautoscript]{revtex4}
\pdfoutput=1
\usepackage{amssymb}
\usepackage{amsmath}
\usepackage{graphicx}
\usepackage{bm}
\usepackage{dcolumn}

\newcommand{\beq}{\begin{equation}}
\newcommand{\enq}{\end{equation}}
\newcommand{\dis}{\displaystyle}

\begin{document}

\title{Transitions of tethered polymer chains: A simulation study with the 
bond fluctuation lattice model}
\author{Jutta Luettmer-Strathmann}
\email[]{jutta@physics.uakron.edu}
\affiliation{
Department of Physics and Department of Chemistry, The University of
Akron, Akron, Ohio 44325-4001} 
\author{Federica Rampf}
\author{Wolfgang Paul}
\author{Kurt Binder}
\affiliation{
Institut f{\"{u}}r Physik, Johannes-Gutenberg-Universit{\"{a}}t,
Staudinger Weg 7, D-55099 Mainz, Germany}
\date{\today}
\begin{abstract}
A polymer chain tethered to a surface may be compact or extended,
adsorbed or desorbed, depending on interactions with the surface and
the surrounding solvent. This leads to a rich phase diagram with a 
variety of transitions. 
To investigate these transitions 
we have performed Monte Carlo simulations of a bond-fluctuation model
with Wang-Landau and umbrella sampling
algorithms in a two-dimensional state space.
The simulations' density of states results have been
evaluated for interaction parameters spanning the range
from good to poor solvent
conditions and from repulsive to strongly attractive surfaces. 
In this work, we describe the simulation method and 
present results for the overall phase behavior and for some of the
transitions. 
For adsorption in good solvent, we compare with Metropolis Monte Carlo
data for the same model and find good agreement between the
results. 
For the collapse transition, which occurs when the 
solvent quality changes from good to poor, we consider two situations 
corresponding to three-dimensional (hard surface) and
two-dimensional (very attractive surface) chain conformations,
respectively. 
For the hard surface, we compare tethered chains with free chains and find
very similar behavior for both types of chains.
For the very attractive surface, we find the two-dimensional chain
collapse to be a two-step transition with the same sequence of transitions
that is observed for three-dimensional chains: 
a coil-globule transition that changes the
overall chain size is followed by a local rearrangement of chain
segments.

\end{abstract}

\maketitle

\section{Introduction}

Chain molecules anchored at surfaces are a part of many physical
systems. Examples include polymer chains grafted to colloidal
particles \cite{na83}, proteins projecting from cell membranes
\cite{ne04}, block copolymers at liquid air interfaces
\cite{ha92b,ke00}, and long chain molecules attached to 
inorganic surfaces for study with atomic force microscopes \cite{ra06}. 
Parameters such as composition of the chains, grafting density, 
and interactions with the surface and the environment 
affect the properties of these systems by determining the
conformations of the chains.
In this work we focus on the effects of solvent quality
and surface interactions and investigate individual polymer chains 
tethered to a flat, impenetrable surface. 
A tethered chain near an attractive surface has many of the
conformation characteristics of a free chain adsorbed from
solution. Likewise, a tethered chain near a hard surface has much in
common with an isolated free chain. These similarities allow us to 
compare with known results for free chains to validate our
approach. They also suggest that our new results are relevant to
the process of surface adsorption from dilute polymer solutions. 
  
As is well known, the collapse transition that a flexible polymer
chain undergoes in dilute solution when the solvent quality
deteriorates is one of the fundamental problems in the statistical
mechanics of polymers. It is of crucial importance in understanding
the phase diagrams of polymer solutions and still offers surprising
insights \cite{pa07}. 
Of similar importance is the adsorption transition in good solvent, 
where the conformations of a tethered chain change from 
``mushroom'' to ``pancake'' configurations as 
the strength of the surface-monomer attraction increases. 
This problem has attracted longstanding intense attention as a 
basic phenomenon of polymer chains interacting with interfaces
\cite{ei93}. 
Of course, both ``transitions'' are sharp thermodynamic
phase transitions only in the (thermodynamic) limit when the chain
length tends to infinity, but understanding how this limit is
precisely approached is one of the challenges here.

In the present work, we go one step beyond the problems outlined
above, by considering the interplay of adsorption and collapse.
This problem is essential for understanding polymeric aggregates
on surfaces when no solvent is present( e.g.\, adsorbed polymers on a 
surface exposed to air!). It is also a first step towards
the treatment of heteropolymers at surfaces, 
a system suited to give insight into the behavior of biopolymers at
biological interfaces, membranes, etc.\,
since these polymers of biological interest often assume rather dense
conformations. 
However, even the case of a homopolymer
at a surface undergoing adsorption competing with collapse is
a very difficult problem, and despite earlier 
numerical studies
\cite{vr96,vr98,vr99,si01,ra02b,mi03,kr05,ow07,ba05,ba06} 
not yet fully understood. 
In particular, the question how the polymer model may affect the
observed phenomena merits further investigation. 
In most of the existing work, the polymer has been modeled as an
interacting self-avoiding walk on a simple cubic lattice (ISAW). In
the ISAW model, 
the bond length is fixed to the size of the lattice constant and the
bond angles are restricted to the values of $\pi/2$ and $\pi$. 
This leads to very specific chain conformations in poor-solvent
conditions, where a chain near an attractive surface fills the volume
of a rectangular box \cite{mi03,kr05,ba06}. 
In contrast, an off-lattice model of a polymer chain under such
conditions has chain conformations whose shapes resemble the spherical 
caps that are typical for partially wetting liquid drops
\cite{me03}.  
In this work, we employ the bond fluctuation (BF) model to
investigate tethered chains. In the BF model, there are five allowed
bond lengths leading to a large set of allowed bond angles 
\cite{ca88b,bi95,la00b}. In terms of chain conformations, the BF model
may be considered an intermediate between the ISAW lattice
model and coarse-grained off-lattice models such as bead-spring models. 
In crystallizable homopolymers such as polyethylene and poly(ethylene
oxide) the local structure of the chains leads to an alignment of
chain sections and lamella formation upon crystallization \cite{ch05}. 
The necessity for mesoscopic conformational ordering accompanying this
alignment gives rise to a large entropic barrier to crystallization
and makes kinetic effects a dominant factor in polymer
crystallization. Therefore it is difficult to separate thermodynamic
effects from non-equilibrium effects. In coarse-grained models of
polymers, such as the BF model and typical bead-spring models, one can
choose force fields with sufficient local flexibility to allow for
crystallization by local rearrangement of monomers. This reduces the
entropic barrier to crystallization and allows the systems to reach
ordered equilibrium states. While their lack of lamella formation
hampers the comparison with experimental data for simple
crystallizable homopolymers, the lower entropic barriers of
coarse-grained models allow us to study equilibrium aspects of polymer
crystallization. 
The BF model has proven to be a very
useful model for a range of polymeric systems (see, for example, 
Refs.~[\onlinecite{ca88b,bi95,la00b,bi97,ba00,bi05}]) but has not yet
been investigated in the present context.

\begin{figure}[htb]
\includegraphics*[width=3.2in]{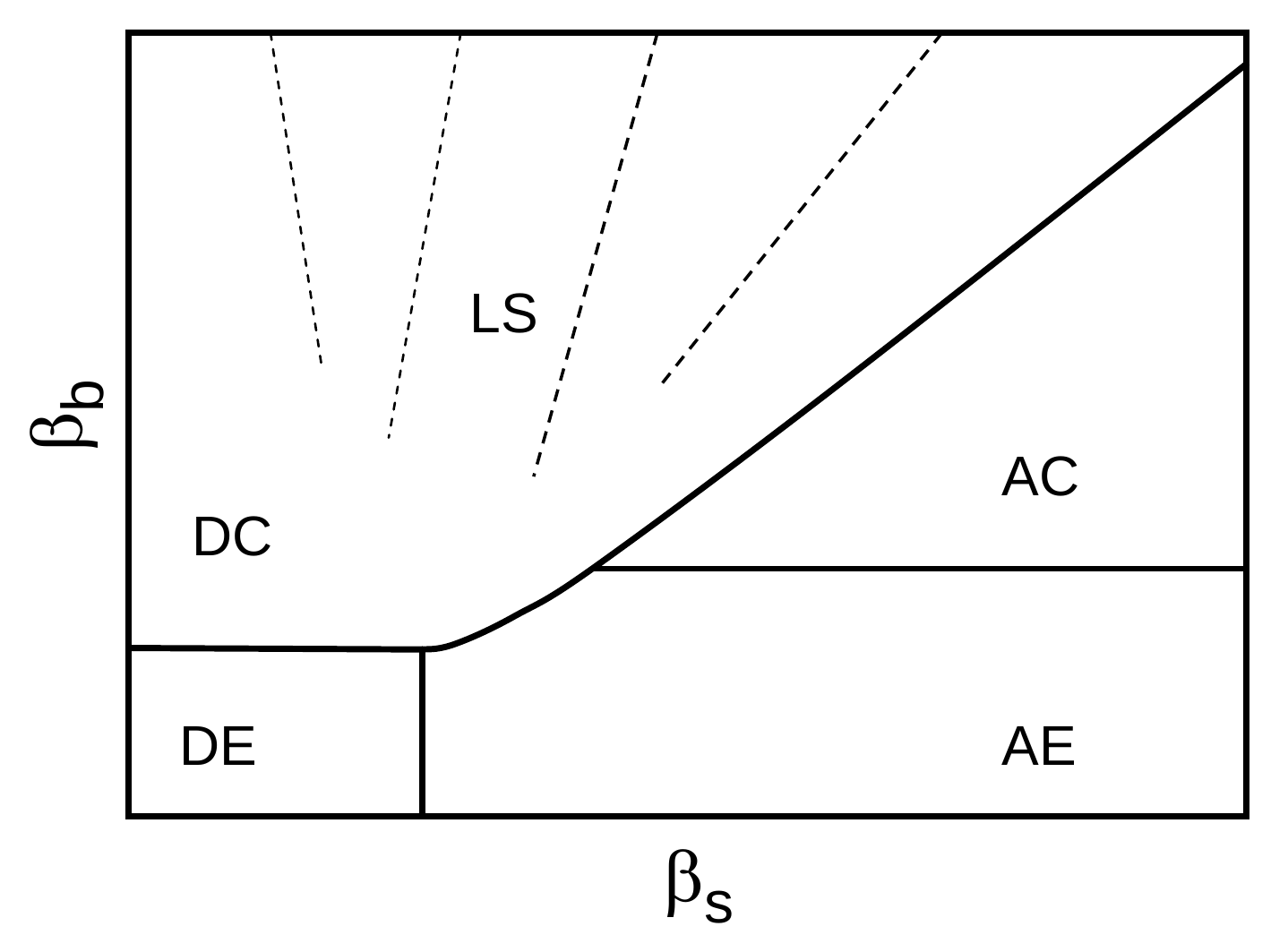}
\caption{\label{fig1}
Schematic phase diagram for tethered chains of finite length in the space of
field variables $\beta_b$ and $\beta_s$ (see
Sec. \protect\ref{eval_dos} for the formal definition). As discussed
in the text, increasing values of $\beta_s$ and $\beta_b$ 
correspond to increasingly attractive
surface and monomer-monomer interactions, respectively. 
The lines indicate transitions between states identified by the 
following abbreviations: DE for desorbed extended (mushroom), AE for
adsorbed extended (pancake), DC for desorbed compact, AC for adsorbed
compact, and LS for layered states. The solid lines indicate
transitions that are expected to become true phase transitions in the
limit of infinite chain length. The dashed lines represent structural
transitions observed for finite size chains only. The ending of the
dashed lines indicates that, in simulations, these structural
transitions can no longer be uniquely identified in regions of field 
parameters, where several transitions compete with each other.
}
\end{figure}

The combined effects of surface interactions and solvent quality lead
to a variety of transitions for a polymer chain tethered to a surface. 
In Fig.~\ref{fig1} we present a schematic phase diagram 
that represents insights from theoretical and simulation work
on lattice models of tethered polymer chains
\cite{vr96,vr98,vr99,si01,ra02b,mi03,kr05}  
including results presented in this work.
The field variables $\beta_s$ and $\beta_b$ 
are a combination of inverse temperature and interaction parameters
\cite{vr96,kr05} and will be defined in
Section~\ref{eval_dos}. The field $\beta_s$ describes the effect of
surface interactions and increases with increasing attraction between
the surface and the chain segments. The field $\beta_b$ describes the
net interactions between monomers of a polymer chain in solvent. 
$\beta_b$ is small for a chain in a good solvent and increases with
decreasing solvent quality, corresponding to 
increasing net attractive interactions between monomers. 

The regions marked DE (desorbed extended) and AE (adsorbed extended)
correspond to good solvent conditions. With field values in the DE
region, a tethered chain assumes extended three dimensional
conformations that are often described as mushrooms. 
An increase of the surface field $\beta_s$ leads to chain adsorption. 
For fields in the AE region, the chain conformations are (nearly) two
dimensional and extended; they are sometimes called pancake
conformations. The transition from DE to AE states has been
investigated with theoretical and 
simulation methods for a variety of models (see, for example,
Refs.~[\onlinecite{ei82,de93,me02,me03,de04}]). 
In the limit of infinite chain length, the adsorption transition
is sharp and corresponds to a multicritical point. 

The regions marked DC (desorbed compact), AC (adsorbed compact), and
LS (layered states) correspond to poor solvent conditions. 
Subject to fields in the DC region, a chain assumes compact
three-dimensional globule conformations and for fields in the AC region
the corresponding two-dimensional conformations. 
The region labeled LS is characterized by a competition between the
effects of attractive surface and monomer interactions.

For weak surface fields $\beta_s$, an increase in the field $\beta_b$
leads to chain collapse, i.e.\ the coil-globule transition from
desorbed extended (DE) to desorbed compact (DC) states. 
For free chains, the coil-globule transition is a continuous
transition which may be followed by a first-order transition that is
associated with spatial ordering of the chain segments (see, for
example, Refs.~[\onlinecite{zh96d,zh97,ra05b}]).  
From theoretical work on the ISAW model, the line describing the
coil-globule transition is expected 
to be a straight line parallel to the $\beta_s$ axis \cite{vr96} and 
perpendicular to the adsorption transition line. 
For strongly attractive surfaces, i.e.\ for large $\beta_s$ values,
an increase in $\beta_b$ leads to a transition from adsorbed extended
(AE) to adsorbed compact (AC) states. In analogy with the coil-globule
transition of a free chain in two dimensions, this transition is
expected to be continuous for ISAW models \cite{vr98}.  

Adsorption in poor solvent occurs for high $\beta_b$ values as the
surface field $\beta_s$ is increased. It takes
a tethered chain from desorbed globule conformations in three
dimensions (DC) to compact, single layer conformations (AC) through
intermediate states (LS) that are not yet fully understood and appear
to be model-dependent.
Singh {\it et al.}~\cite{si01,ra02b,mi03} investigated {\it
  untethered} chains by exact 
enumerations of interacting self-avoiding walks on a simple cubic
lattice (ISAW  model). In poor solvent conditions, a free 
chain assumes its most compact conformation which, in the ISAW model,
has the shape of a cube with side length $l\propto N^{1/3}$, where $N$
is the length of the chain. In the presence of a (slightly) attractive
surface, the cubic globule attaches itself to the surface and makes a
number of surface contacts proportional to the area of a face of the
cube, i.e.\ $\propto N^{2/3}$. 
Singh {\it et al.}~\cite{si01,ra02b,mi03} interpret the transition
from the detached to the attached globule as a surface transition from
the DC phase to a ``surface attached globule'' phase, which occupies
the LS region in Fig.~\ref{fig1}, and is followed by a second phase
transition at high $\beta_s$ values to the AC phase. 
For {\it tethered} chains in the ISAW model, 
theoretical arguments suggest that there is only one 
true phase transition (non-analyticity of the free energy in
the limit of infinite chain length) 
as $\beta_s$ is varied at constant $\beta_b$ \cite{vr96}. 
More recent simulations of tethered chains in the ISAW model by
Krawczyk {\it et al.}~\cite{kr05} found the chains to undergo a series
of layering transformations in the LS region corresponding to
transitions between compact chain conformations with different numbers
of layers parallel to the surface. This work identified scaling
relations which suggest that, with increasing chain length, the
transitions become sharper and more numerous, and that the transitions
merge into a single transition in the limit of infinite chain length. 
In the diagram of Fig.~\ref{fig1}, we represent the true phase
transition by a solid line between the AC and LS regions. The dashed
lines in Fig.~\ref{fig1} indicate structural/layering transitions that 
may not become phase transitions in the infinite
chain limit. These transitions are important, however, 
since for many problems the behavior of 
finite-length chains is of physical interest. 

In this work, we investigate transitions of tethered polymer chains
with the aid of density-of-states simulations of a bond-fluctuation
model.
In the work presented here we focus on characteristic
features of adsorption in good
solvent (DE to AE) and chain collapse in two and three dimensions (DE
to DC and AE to AC). We also construct a phase portrait for our 
longest chain. In a forthcoming publication we shall  
discuss adsorption in poor solvent and investigate some 
states and transitions in detail. 
In this article, we introduce the bond-fluctuation model for a
tethered chain and describe the simulation
methods in Sec.~\ref{method} and the data evaluation in
Section~\ref{eval_dos}. 
Results of our simulations are presented and discussed in 
Sec.~\ref{results}, followed by a brief summary and conclusions in
Sec.~\ref{summary}. 
Technical details of the simulations undertaken for this work are
  presented in Appendix \ref{details}.

\section{Model and simulation method}\label{method}

In this section, we present the model for a surface-attached
polymer chain and the simulation methods employed in this work. 
Details of the simulation protocol that are of interest to those
who want to reproduce or extend the work but that are not required for 
an understanding of the following sections have been relegated to 
Appendix \ref{details}.

In the bond fluctuation (BF) model \cite{ca88b,bi95,la00b}, monomers of a
polymer chain occupy 
sites on a simple cubic lattice. The bond lengths between
monomers are allowed to vary between $b = 2a$ and $b = \sqrt{10}a$,
where $a$ is the lattice constant, which we set to unity, $a=1$. 
A tethered chain is represented by a chain whose first monomer is
fixed just above a hard surface. The position of the surface is the
$x$-$y$ plane at position $z=0$ in a Cartesian coordinate system; the
coordinates of the fixed monomer are (1,1,1). 
Monomers at $z=1$ are considered to be in contact with the surface and
contribute an amount $\epsilon_s$ to the energy. 
The interactions between monomers depend on the distance. 
A pair of monomers $i$ and $j$ at a distance $r_{ij}$ 
contributes an amount of
$\epsilon_b$ to the energy when $4 \leq r_{ij}^2 \leq 6$. 
Distances $r_{ij}^2 < 4$ are prohibited by hard core repulsion, 
while monomers do not interact for $r_{ij}^2 > 6$. 
The total energy of the system is given by
\begin{equation}\label{energy}
E(n_s,n_b) = n_s\epsilon_s + n_b\epsilon_b ,
\end{equation}
where $n_s$ and $n_b$ are the number of monomer-surface and
monomer-monomer contacts, respectively. We refer to a pair of contact
numbers, $(n_s,n_b)$, as a state of the system.

In the Monte Carlo simulations described here, two types of elementary
moves were carried out; local moves that displace a monomer to a
nearest neighbor lattice site and pivot moves about the $z$-axis. 
Simulations were performed for chains of length $N = 16$, $N = 32$,
and $N = 64$. In all simulations, $N$ local move attempts were
followed by ten pivot move attempts. We refer to this sequence as one
Monte Carlo (MC) step. 

\subsection{Wang-Landau algorithm for a 2-dimensional state space}\label{WL}

The density of states (dos), $g(n_s,n_b)$ is the number of
configurations for a given state $(n_s,n_b)$ of the system. 
The Wang-Landau (WL) algorithm \cite{wa01b,la04b} is an iterative
Monte Carlo simulation 
method for constructing the density of states. Originally formulated
for a one-dimensional state space, characterized by the total energy,
it has been extended to state spaces of higher dimensions
\cite{zh05,ma05,tr05}. In this work, the state space is 
the two dimensional space of contact numbers $n_s$ and $n_b$. 

In the WL algorithm, an elementary move attempt from a 
state $(n_s,n_b)$ to a state $(n'_s,n'_b)$ is accepted with probability 
\begin{equation}\label{accpt}
p\left( (n_s,n_b) \rightarrow (n'_s,n'_b) \right) = 
\min\left(\frac{g(n_s,n_b)}{g(n'_s,n'_b)},1\right),  
\end{equation}
where $g(n_s,n_b)$ is the current estimate for the density of states. 

The initial guess for the density of states is $g(n_s,n_b)=1$ for all
states $(n_s,n_b)$.   
At each iteration level, elementary move attempts are followed by
an update of the density of states and the histogram of visits to the
states, $h(n_s,n_b)$.  After an attempted move from a
state $(n_s,n_b)$ to a state $(n'_s,n'_b)$ the updates are given by 
\begin{eqnarray}\label{update}
\mbox{\hspace{-0.5em}} \text{if accepted:} \mbox{\hspace{-0.8em}}&&
\left\{
\begin{array}{rcl}
\ln(g(n'_s,n'_b))&\rightarrow&\ln(g(n'_s,n'_b))+\ln(f) , \\
h(n'_s,n'_b)&\rightarrow&h(n'_s,n'_b)+1 ,
\end{array}
\right. \\
\mbox{\hspace{-0.5em}}\text{if rejected:} \mbox{\hspace{-0.8em}}&&
\left\{
\begin{array}{rcl}
\ln(g(n_s,n_b))&\rightarrow&\ln(g(n_s,n_b))+\ln(f) , \\
h(n_s,n_b)&\rightarrow&h(n_s,n_b)+1 , \label{updateb}
\end{array}
\right. 
\end{eqnarray}
where $f$, with $f>1$, is the refinement factor.  
An iteration is considered complete when the histogram satisfies a
flatness criterion. In a typical Wang-Landau simulation
\cite{wa01b,la04b}, a histogram is  
considered flat when the smallest value in the histogram is at
least 80\% of the average value of the histogram entries.
At that point, the histogram is reset to $0$ for
all states and the refinement factor is reduced 
before the next iteration is started. 
The adjustments to the density of states become smaller with each
iteration level; we used refinement levels with 
$\ln(f_{k})=2^{-(k-1)}$ for $k\in\{2,20\}$.

\begin{figure}[htb]
\includegraphics*[width=3.2in]{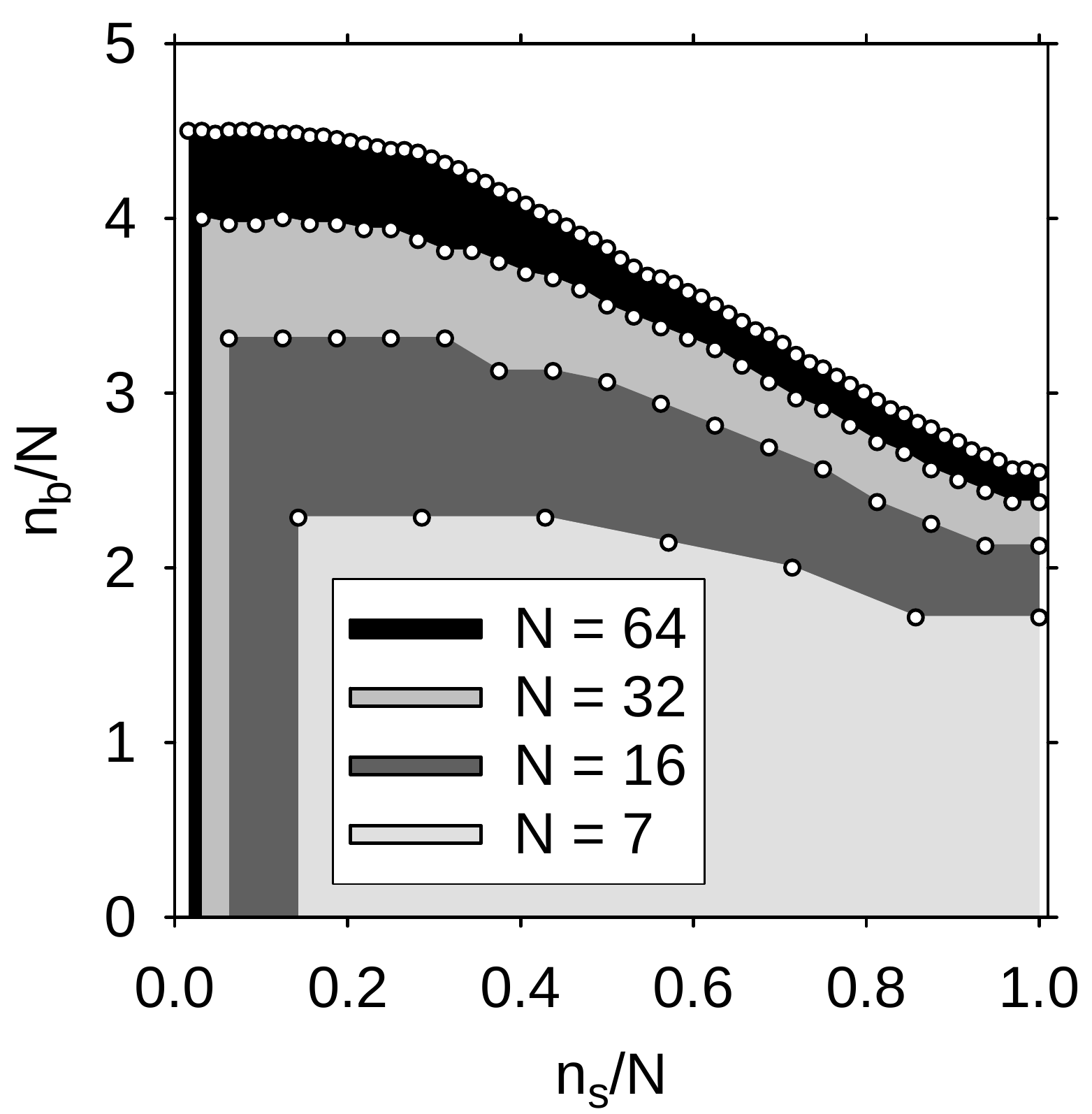}
\caption{\label{dos_range}
Ranges of accessible states for tethered chains of lengths $N=7$,
16, 32, and 64. The shaded areas approximate the ranges, the
symbols represent the realized states with the 
highest number of bead contacts $n_b$ for given number of surface
contacts $n_s$. The results for $N=7$ were obtained in exact
enumeration.}
\end{figure}

For tethered chains, the range of accessible states $(n_s,n_b)$ is not
known a priori. Since the maximum number
of surface contacts $n_s$ is equal to the number of monomers $N$, the
maximum value of the ratio $n_s/N$ is unity for all
chain lengths. However, the maximum value of $n_b/N$, the number of
monomer-monomer contacts per monomer, increases with chain length 
as illustrated in Fig.~\ref{dos_range}. 
In addition, for each chain length, the maximum number
of monomer-monomer contacts decreases with increasing number of
surface contacts.
In the work presented here, we allow the range of considered
states to change during the simulations. 
During the Wang-Landau simulations, 
when a configuration to a previously unvisited state $(n'_b,n'_s)$ 
appears, the new state is assigned the initial dos value
$g(n'_s,n'_b)=1$ and the move is accepted or rejected with the usual
criterion.  

For simulations of free chains in the bond fluctuation model, it was
found that updating after accepted moves only may lead to shorter
simulation times without affecting the results for the density of
states \cite{ra06b}. 
In the work presented here, we carried out simulations where either
both updates, Eqs.~(\ref{update}) and (\ref{updateb}), or only the
updates after accepted moves, Eq.~(\ref{update}), were carried out. 

\subsection{Multiple replica algorithm}\label{multiple}

In a multiple replica simulation, the state space is divided into a
number of overlapping regions. Standard simulation steps are carried
out separately in each replica. After each MC step, 
an attempt is made to exchange the chains of two neighboring
replicas if their configurations belong to states in the overlap
region of the replicas. The acceptance criterion for such a
replica-swapping move is 
\begin{equation}\label{swap_accpt}
\begin{array}{l}
p\left( [(n_s,n_b)_i, (n'_s,n'_b)_{i+1}] 
\rightarrow (n'_s,n'_b)_i, (n_s,n_b)_{i+1}] 
\right)  \\
\mbox{\hspace{1em}} 
= \min\left(\frac{\dis g(n_s,n_b)_{i+1}}{\dis g(n'_s,n'_b)_i}
\frac{\dis g(n'_s,n'_b)_{i+1}}{\dis g(n_s,n_b)_i},1\right),
\end{array}
\end{equation}
where $i$ and $i+1$ are the indices for neighboring replicas.
When using the method to generate the density of states, the dos
values are updated after each attempted move in the usual way, see
Eqs.~(\ref{update}) and (\ref{updateb}).
The density of states for the whole state space is obtained at the end
of the simulation by combining the results for each 
replica.
For the results presented here, we used
regions that were defined  by intervals in the number of bead-bead
contacts $n_b$ and contained all possible surface contacts. 
We experimented with regions of different size and different ranges of
overlap and found that small but systematic errors in the density
of states occurred unless each state was covered by the same number of
replicas. We worked with a total of three to five overlapping replicas
covering each region with two replicas.

\subsection{Global update algorithm}\label{global}

When the WL algorithm is applied to large two-dimensional
state spaces much time is spent accumulating the large $\ln(g)$
values belonging to the interior of the state space. 
Introducing global updates \cite{zh05} improves the
efficiency by allowing the simulation to spend more
time exploring the edges of the current state space. 
In a global update algorithm, a simulation is started at a given
refinement level 
and the current values of $\ln(g(n_s,n_b))$ are compared with a
threshold value $\omega$. Once 
density of states values above the threshold, $\ln(g(n_s,n_b)) >
\omega$ are found, they are augmented in the following way:
\begin{equation}\label{shift}
\ln(g) \rightarrow 
\ln(g) + \kappa\exp
\left(\frac{\displaystyle - \lambda}{\displaystyle \ln(g)-\omega}\right) 
\Theta(\ln(g)-\omega) ,
\end{equation}
where $\Theta$ is the Heaviside step function. 
The exponential function dampens the shift from a maximum value
of $\kappa$ for states with $\ln(g)$ values well above the threshold
to zero for states at the threshold; the parameter $\lambda$ determines
the range over which the shift is phased out.  
The simulation then continues with standard steps and local updates to
the density of states.
Immediately after a global shift, states that
were outside the shifted region are preferentially sampled. With time,
their $\ln(g)$ values increase thereby decreasing the acceptance
rates for moves to those states and leading to uniform sampling over
previously shifted and unshifted states. 
Once uniform growth over the current state space has resumed 
the next global update is considered. This continues until a flatness
criterion is satisfied or the simulation is terminated by hand.

To determine when uniform growth has been achieved 
in our simulations, we monitor the difference 
$\Delta \ln(g)$ between the current value of
the log-density of states and that at the time of the global shift. 
When for each previously shifted state the difference $\Delta \ln(g)$  
is larger than a fixed parameter $\rho$, 
we conclude that uniform growth has been achieved.
We found that the values for the parameters $\kappa$,
$\lambda$, $\omega$ and $\rho$ can have a large effect on the quality
of the results. 
In particular, systematic deviations in the results for the density of
states occur when the global updates are too large and occur too
frequently. 

\subsection{Umbrella sampling}\label{umbrella}

In multicanonical or umbrella sampling simulations with single
histogram reweighing, a good estimate for the density of states
$g(n_s,n_b)$ is  required as input (see, for example,
Refs. [\onlinecite{be92,la00b,mi01c,ja03b}]). 
We performed umbrella sampling simulations with a WL result as the
initial dos for all simulation results reported here. 
During the simulation,  
elementary moves are accepted with the usual criterion, see
Eq.~(\ref{accpt}), and the histogram is updated after accepted and
rejected moves. The dos values, however, are updated only at the end 
of the simulations. 
In a typical single histogram reweighing step, the logarithm of the
final histogram entry $h(n_s,n_b,t_f)$,
is added to the original log-density of states
values,
\begin{equation}\label{reweigh}
\ln(g'(n_s,n_b)) = \ln(g(n_s,n_b)) + \ln(h(n_s,n_b,t_f)).
\end{equation} 
In the work presented here, we record cumulative histograms at regular
intervals, determine the slope $m(n_s,n_b)$ of the histogram entries
as a function of time, and estimate the final histogram entry
from the slope $h(n_s,n_b,t_f)\simeq m(n_s,n_b)t_f$. This gives a
slightly more reliable dos estimate for states that are not visited in 
every block.

\subsection{Metropolis algorithm}\label{metropolis}

For comparison with results from the density of states algorithms
and in order to sample efficiently some parts of phase space during
the production stage, some simulations were performed with a
Metropolis acceptance criterion at fixed fields $\beta_s$ and
$\beta_b$. The probability for accepting a move from a 
state $(n_s,n_b)$ to a state $(n'_s,n'_b)$ 
in the Metropolis algorithm may be written as 
\begin{equation}\label{met_accpt}
\begin{array}{l}
p\left( (n_s,n_b) \rightarrow (n'_s,n'_b) \right) \\
\mbox{\hspace{1em}} = 
\min\left(\exp[\beta_s(n'_s-n_s)+\beta_b(n'_b-n_b)],1\right).
\end{array}
\end{equation}

\subsection{Production stage}\label{production}

In a production simulation, chain conformations are evaluated at
regular intervals to accumulate configurational properties as a
function of the pairs of contact numbers $(n_s,n_b)$. 
We determined a range of configurational properties including chain
dimensions and density profiles. Furthermore, configurations were
stored for a detailed analysis of the chain structure. 
In this work, we present results for the squared bond length 
$B^2=\sum_{i}(\mathbf{r}_{i+1}-\mathbf{r}_i)^2/(N-1)$, 
where $\mathbf{r}_i$ is the position vector of monomer $i$, and for
the radius of gyration $R_g$. 
In order to investigate the effect of the surface, we
calculate parallel and perpendicular contributions to $R_g^2$
according to 
\begin{eqnarray}\label{Rg2z}
R_{g,z}^2 &=& \frac{1}{N^2}\sum_{i<j}(z_i-z_j)^2, \\
R_{g,xy}^2 &=&
\frac{1}{N^2}\sum_{i<j}\left[(x_i-x_j)^2+(y_i-y_j)^2\right], \label{Rg2xy}
\end{eqnarray}
where $i,j\in\{1,\ldots,N\}$ and where $x_i,y_i,z_i$ are the Cartesian
coordinates of monomer $i$, and $R_g^2=R_{g,z}^2+R_{g,xy}^2$.

When production simulations are performed with the acceptance
criterion of the Wang-Landau algorithm all states $(n_s,n_b)$ are
visited approximately equal numbers of times. 
This may lead to insufficient sampling of states $(n_s,n_b)$ with a 
diverse set of chain conformations. 
In order to improve the statistics for the configurational properties
belonging to such states $(n_s,n_b)$, we performed 
additional production simulations with a Metropolis Monte Carlo
algorithm for fields between $\beta_s = \beta_b = 0$ and $\beta_s =
\beta_b = -2$.  
In these simulations, the visitation histograms are approximately
Gaussian with maxima at the most probable states for these fields.

For all chain lengths, we performed more than one type of production
simulation. During production, 
block averages \cite{ne99,la00b} were calculated and the statistical
uncertainty of an average value was  
determined from the standard deviation of the block averages.
Results from different production simulations were combined as
weighted averages, where the weights were based on the uncertainty
estimates obtained for each quantity in the individual
production simulations.

\section{Evaluation of the density of states}\label{eval_dos}

The canonical partition function for a tethered chain is given by  
\begin{equation}\label{zpart}
Z = \sum_{n_s.n_b}g(n_s,n_b)e^{-\beta E(n_s,n_b)}
\end{equation}
where $\beta=1/k_\mathrm{B}T$, $T$ is the temperature, 
$k_\mathrm{B}$ is Boltzmann's constant, and 
$g(n_s,n_b)$ is the density of states. 
The energy of the state, $E(n_s,n_b) = n_s\epsilon_s + n_b\epsilon_b$,
depends on the contact
numbers $n_s$ and $n_b$ and the interaction parameters 
$\epsilon_s$ and $\epsilon_b$. 
In this work, we employ combinations of the interaction parameters and
the inverse temperature as field variables \cite{vr96,kr05} 
\begin{equation}\label{fields}
\beta_s = -\epsilon_s\beta , \mbox{\hspace{3em}}
\beta_b = -\epsilon_b\beta ,
\end{equation}
and write the partition function as 
\begin{equation}\label{zpartf}
Z(\beta_s,\beta_b) = \sum_{n_s,n_b}g(n_s,n_b)e^{\beta_s n_s
}e^{\beta_b n_b}. 
\end{equation}
For fixed fields $\beta_s$ and $\beta_b$,  
the probability $P(n_s,n_b;\beta_s,\beta_b)$ for 
states with contact numbers $(n_s,n_b)$ is given by 
\begin{equation}\label{prob}
P(n_s,n_b;\beta_s,\beta_b) =
\frac{1}{Z}g(n_s,n_b)e^{\beta_s n_s}e^{\beta_b n_b} ,
\end{equation}
and the average values of quantities $Q(n_s,n_b)$ 
at given fields $\beta_s$ and $\beta_b$ are calculated from 
\begin{equation}\label{average}
\langle Q \rangle = \sum_{n_s,n_b}Q(n_s,n_b)P(n_s,n_b;\beta_s,\beta_b)
\equiv Q(\beta_s,\beta_b).
\end{equation}
At the end of this section, we describe how we estimate
uncertainties for the averages determined in this way.

With the definition of a free energy 
\begin{equation}\label{free}
G(\beta_s,\beta_b) = -\ln(Z) ,
\end{equation}
the average number of surface contacts $\langle n_s\rangle$ and the 
average number of monomer contacts $\langle n_b\rangle$ are conjugate 
to the field variables $\beta_s$ and $\beta_b$, respectively, 
\begin{eqnarray}\label{ns}
\langle n_s\rangle &=& -\left(\frac{\partial G}{\partial
  \beta_s}\right)_{\beta_b} , \\
\langle n_b\rangle &=& -\left(\frac{\partial G}{\partial
  \beta_b}\right)_{\beta_s}. \label{nb}
\end{eqnarray}
The second derivatives of the free energy with respect to the fields
define susceptibilities $\chi_{\mu\nu}$, $\mu,\nu\in\{s,b\}$ that are
related to the fluctuations in the number of surface and monomer
contacts  
\begin{eqnarray}\label{flucs}
\chi_{ss} &=& -\left(\frac{\partial^2 G}{\partial
  \beta_s^2}\right)_{\beta_b} = \langle n_s^2\rangle - 
\langle n_s \rangle^2 , \\
\chi_{bb} &=& -\left(\frac{\partial^2 G}{\partial
  \beta_b^2}\right)_{\beta_s} = \langle n_b^2\rangle - 
\langle n_b \rangle^2 , \label{flucb} \\
\chi_{sb} &=& -\left(\frac{\partial^2 G}{\partial
  \beta_s \beta_b}\right) = \langle n_s n_b\rangle - 
\langle n_s \rangle \langle n_b \rangle \label{flucsb}
\end{eqnarray}
In terms of these susceptibilities, the heat capacity of the system 
is given by 
\begin{equation}\label{heatcap}
C = \beta^2\left(\langle E^2\rangle - \langle E \rangle^2\right) = 
\beta_s^2\chi_{ss} + \beta_b^2\chi_{bb} + 2\beta_s\beta_b\chi_{sb}. 
\end{equation}

There are alternative methods to evaluate the density of states which
we would like to discuss briefly. The density of states $g(n_s,n_b)$
provides the entropy as a function of the contact values up to an
arbitrary constant $S_0$
\beq\label{entropy}
S(n_s,n_b) = 
k_{\mathrm{B}} \ln(g(n_s,n_b)) + S_0 .
\enq
Starting from the entropy as the thermodynamic potential, 
the fields and susceptibilities may be expressed in terms of
derivatives of the log density of states. For example, 
the fields
may be obtained as $\beta_s(n_s,n_b) = -(\partial\ln(g)/\partial
n_s)_{n_b}$ and $\beta_b(n_s,n_b) = -(\partial\ln(g)/\partial
n_b)_{n_s}$. 
In the thermodynamic limit, $N\rightarrow\infty$, results from different
statistical ensembles are equivalent, however, 
for our chain lengths finite size effects may play a role.
Since an evaluation
of the entropy is hampered by the need to take numerical derivatives of
discrete variables, we employ the free energy
$G(\beta_s,\beta_b)$ unless otherwise indicated. 

In some cases, it is convenient to work with  
a thermodynamic potential that is a function of 
the number of surface contacts $n_s$ and 
the bead-contact field $\beta_b$. 
This free energy $A(n_s,\beta_b)$ may be considered a
Legendre transform \cite{ca85b} of either the entropy $S$ or the free
energy $G$ and may be written as $A(n_s,\beta_b) = -\ln(Z_A)$ ,
with 
\beq\label{zparta}
Z_A(n_s,\beta_b) = \sum_{n_b}g(n_s,n_b)e^{\beta_b n_b}. 
\end{equation}
In this case, the probability for a state with $n_b$ bead contacts is
given by 
\beq\label{proba}
P(n_b;n_s,\beta_b) = \frac{1}{Z_A}g(n_s,n_b)e^{\beta_b n_b}, 
\enq
which allows us to calculate quantities for given $n_s$ and $\beta_b$
without taking numerical derivatives. For example, the average number
of bead contacts is determined from 
\beq\label{nb_A}
\langle n_b(n_s,\beta_b) \rangle = \sum_{n_b} n_b P(n_b;n_s,\beta_b) .
\enq

In general, there are two sources of error when average 
quantities $\langle Q \rangle = Q(\beta_s,\beta_b)$ for given
field values $\beta_s$ and $\beta_b$ are obtained from the contact
number dependent values $Q(n_s,n_b)$ with the aid of
Eq.~(\ref{average}). 
The first source of error is the uncertainty in the density of states,
this is the only uncertainty for quantities such as the
susceptibilities, see Eqs.~(\ref{flucs}) -- (\ref{flucsb}). 
To estimate this uncertainty, we perform several simulations for the
same chain length and calculate the mean value 
$\overline{\ln(g(n_s,n_b))}$ and the standard deviation 
$\sigma_{\ln(g)}(n_s,n_b)$ for each state $(n_s,n_b)$.
The second source of error is the uncertainty
in production results, which we obtain from block averages. 
For properties such as the chain
dimensions, see Eqs.~(\ref{Rg2z})--(\ref{Rg2xy}), this is typically
the larger source of error. 

In order to estimate the effect of the uncertainties in the density of
states on calculated properties $Q$, we generate five synthetic
densities of 
states $\ln(g_i)$ by drawing randomly from Gaussian
distributions centered on $\overline{\ln(g(n_s,n_b))}$ with standard
deviation $\sigma_{\ln(g)}(n_s,n_b)$ for each state $(n_s,n_b)$.
Average values $\langle Q\rangle_i$ are calculated from each 
synthetic density of states $i$ and the mean $\bar{Q}$ and standard
deviations $\sigma_{Q,g}$ of the results are determined. 
The uncertainties of production data are propagated through
Eq.~(\ref{average}), evaluated with $\overline{\ln(g)}$, to
obtain the second contribution to the uncertainty 
$\sigma_{Q,p}$.  
The combined uncertainty is estimated from 
$\sigma_Q = \sqrt{ \sigma_{Q,g}^2 + \sigma_{Q,p}^2}$.

The uncertainty estimates, $\sigma_{Q,g}$ for results that derive
their errors only from the density of states and 
$\sigma_{Q}$ for results that involve production data, 
are shown as error bars in the figures of this work. 
Error bars are omitted when they are smaller than the symbol size or
the line thickness.

\section{Results and discussion}\label{results}

\subsection{Density of states}\label{dos_results}

Density-of-states results were obtained for chains of length 
$N=16$, $N=32$, and $N=64$ with the simulations described in Section
\ref{method}.
Figure~\ref{dos_range} shows the ranges of accessible states for
these chains and also includes results from an exact enumeration for
chains of length $N=7$. 
In each case, the maximum number of bead-bead
contacts decreases with increasing number of surface contacts since  
chain conformations cannot maximize surface and bead contacts
simultaneously. This leads to a competition between bead and surface
contacts at high field values which increases with increasing chain
length.

\begin{figure}[htb]
\includegraphics*[width=3.2in]{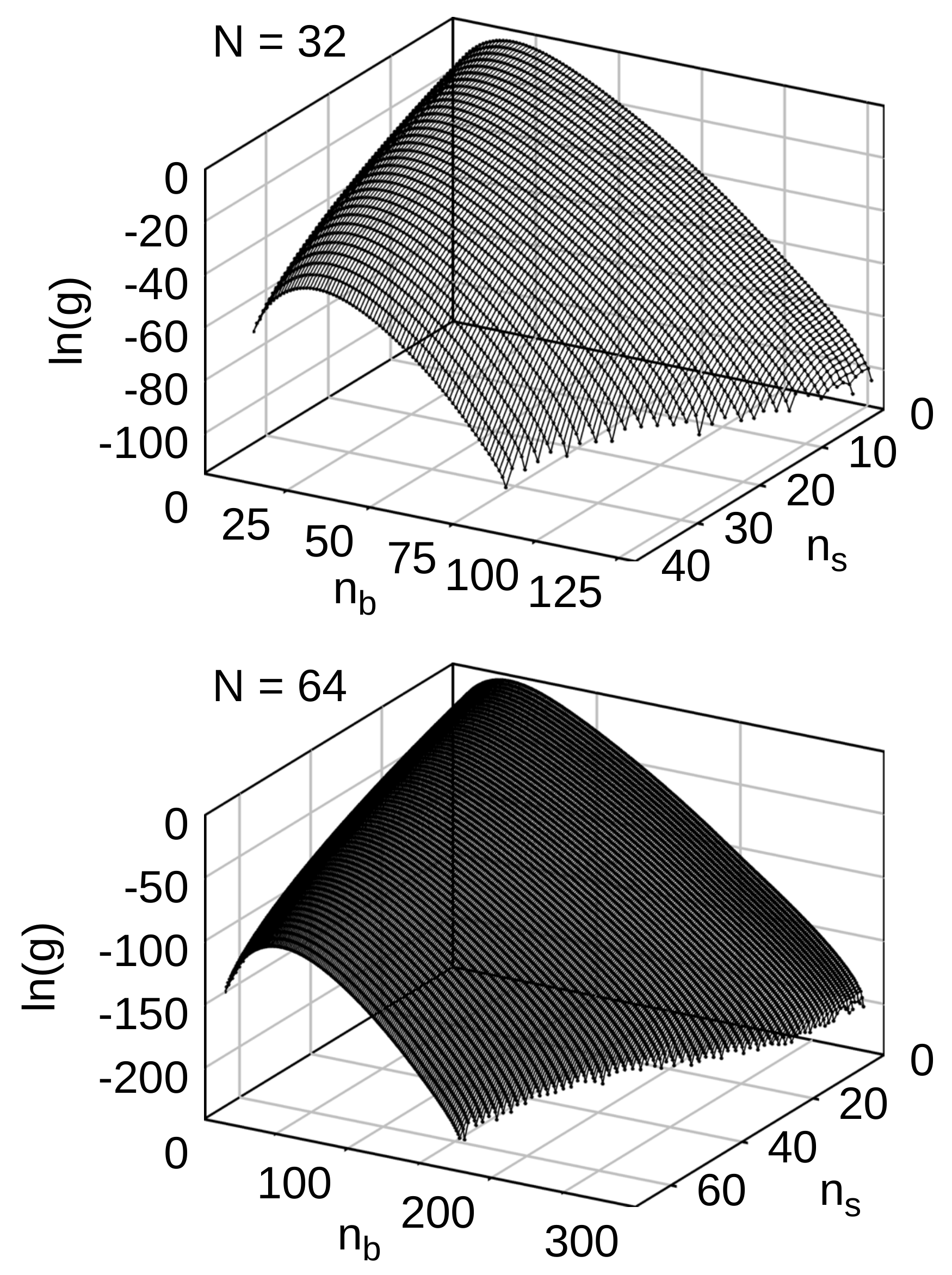}
\caption{\label{dos_plots}
Density of states for tethered chains of length $N=32$ (top) and 
$N=64$ (bottom). 
The surfaces (small symbols connected by straight line segments) 
represent the log
density of states values, $\ln(g)$, as a function of surface contacts,
$n_s$, and bead-bead contacts, $n_b$.}
\end{figure}

\begin{figure}[htb]
\includegraphics*[width=3.2in]{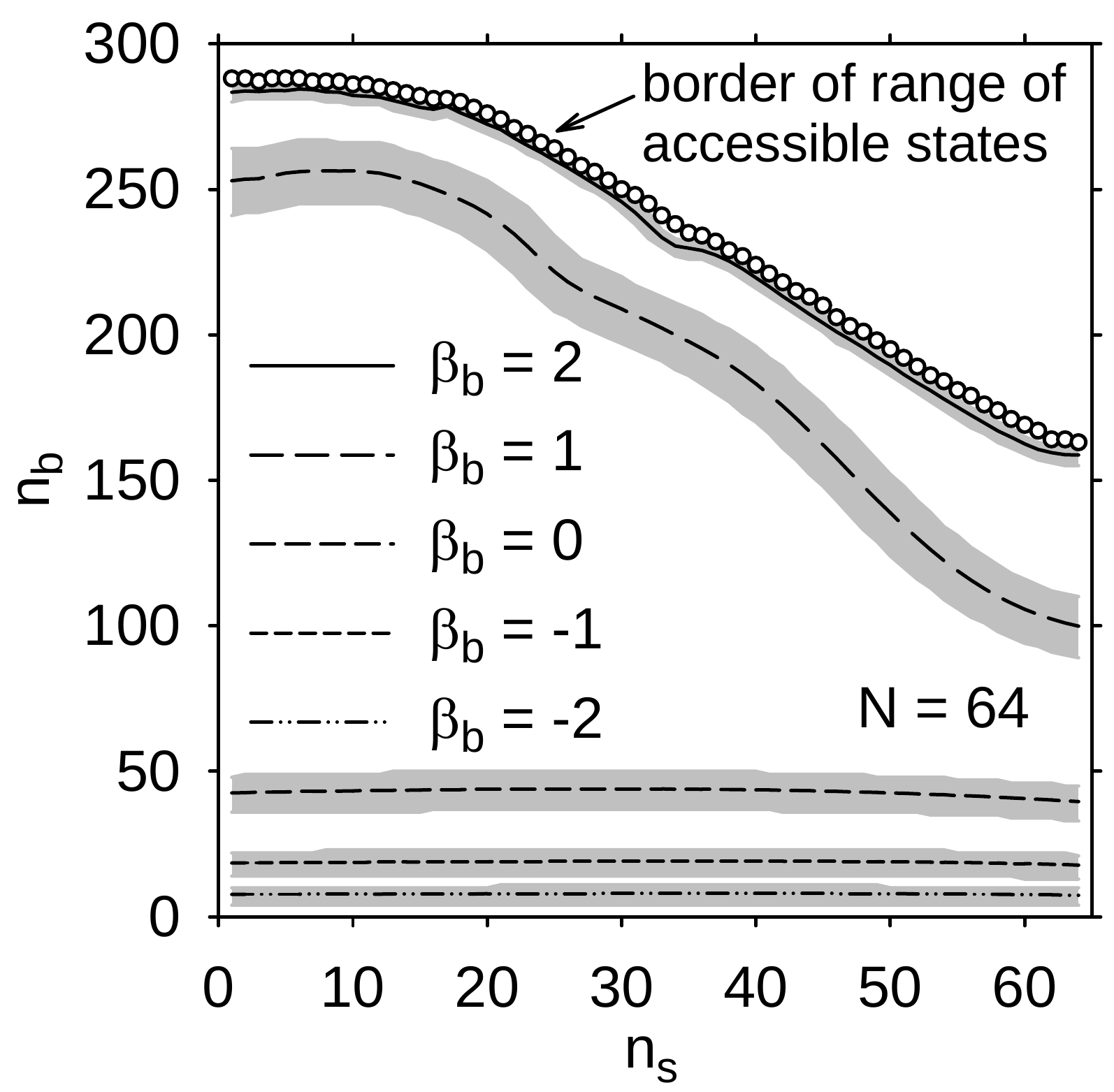}
\caption{\label{pcontours}
Most probable states for given number of surface contacts $n_s$ and
field $\beta_b$. The lines represent the average number of bead contacts 
$\langle n_b(n_s,\beta_b) \rangle$ as function of the number of
surface contacts for five fields $\beta_b$, as indicated in the
figure. 
The shaded areas surrounding the lines indicate the states that
have significant 
probability of occupation; the sum of the probabilities
associated with these states is $2/3$. The symbols at the upper boundary
of the graph indicate the maximum values of $n_b$ for given $n_s$.
}
\end{figure}

In Fig.~\ref{dos_plots} we present results for the log density of
states for chains of length $N=32$ and $N=64$ (the results for $N=16$
show the same qualitative behavior); some characteristic 
numerical values are presented in Table~\ref{table_dos}. 
For each surface
contact value, $n_s$, and for small contact values $n_b$, the  
density of states increases with $n_b$. As $n_b$ increases
further, $\ln(g)$ passes through a maximum and then decreases with
increasing $n_b$. As discussed at the end of Sec.~\ref{eval_dos}, the
slope of $\ln(g(n_s,n_b))$ at fixed $n_s$ is related to the
expectation value of the field $\beta_b(n_s,n_b)$. 
This suggests that states with small bead contact values are
predominantly populated for negative $\beta_b$ fields 
(repulsive interactions between the beads) while
states with large bead contact values are populated for positive 
$\beta_b$ fields (attractive bead-bead interactions).
This is confirmed by an evaluation of the average bead contact number
according to Eq.~(\ref{nb_A}). In Fig.~\ref{pcontours} we present 
results for the average number of bead contacts 
$\langle n_b(n_s,\beta_b) \rangle$ as function of the number of
surface contacts for five fields $\beta_b$. 
For each $n_s$ and $\beta_b$ we also evaluated cumulative
probabilities $\sum_{n_b=0}^{n*_b} P(n_s,\beta_b;n_b)$ and determined
the bead contact numbers $n*_{b,1}$ and $n*_{b,2}$ 
where the cumulative probability first exceeds $1/6$ and $5/6$,
respectively. The states between $n*_{b,1}$ and $n*_{b,2}$
have significant probability of occupation and are indicated by
the shaded areas in Fig.~\ref{pcontours}. 
As expected, the range of significantly populated states shifts from
low-$n_b$ to high-$n_b$ values with increasing field $\beta_b$. For
the largest field shown, $\beta_b = 2$, states on the upper rim of the
range of accessible states start to become populated. Since the
density of states values for rim states have a relatively large
uncertainty, we restrict ourselves in this work to fields $\beta_b
\leq 2$. 
The width of the shaded areas in Fig.~\ref{pcontours} is an indication
for the size of the fluctuations about the mean value. These
fluctuations are larger for $\beta_b=1$ than for the other fields,
since $\beta_b=1$ is closer to the field value of the
coil-globule transition (see Sec.~\ref{collapse}) 
than the others. Geometrically, the size of
the fluctuations is associated with the inverse of the curvature of
the log-density of states surface.

For most accessible states, the density of states is a monotonously
decreasing function of  
$n_s$ at constant $n_b$, i.e.\ the states are predominantly 
occupied for attractive surface interactions. 
Only for a range of states with small $n_s$ and very large $n_b$
values is the slope of $\ln(g)$ for fixed $n_b$ positive. These states
become populated predominantly when $\beta_s$ is negative and
$\beta_b$ is large and positive.

\subsection{Adsorption in good solvent}\label{adsorption}

\begin{figure}[htb]
\includegraphics*[width=3.2in]{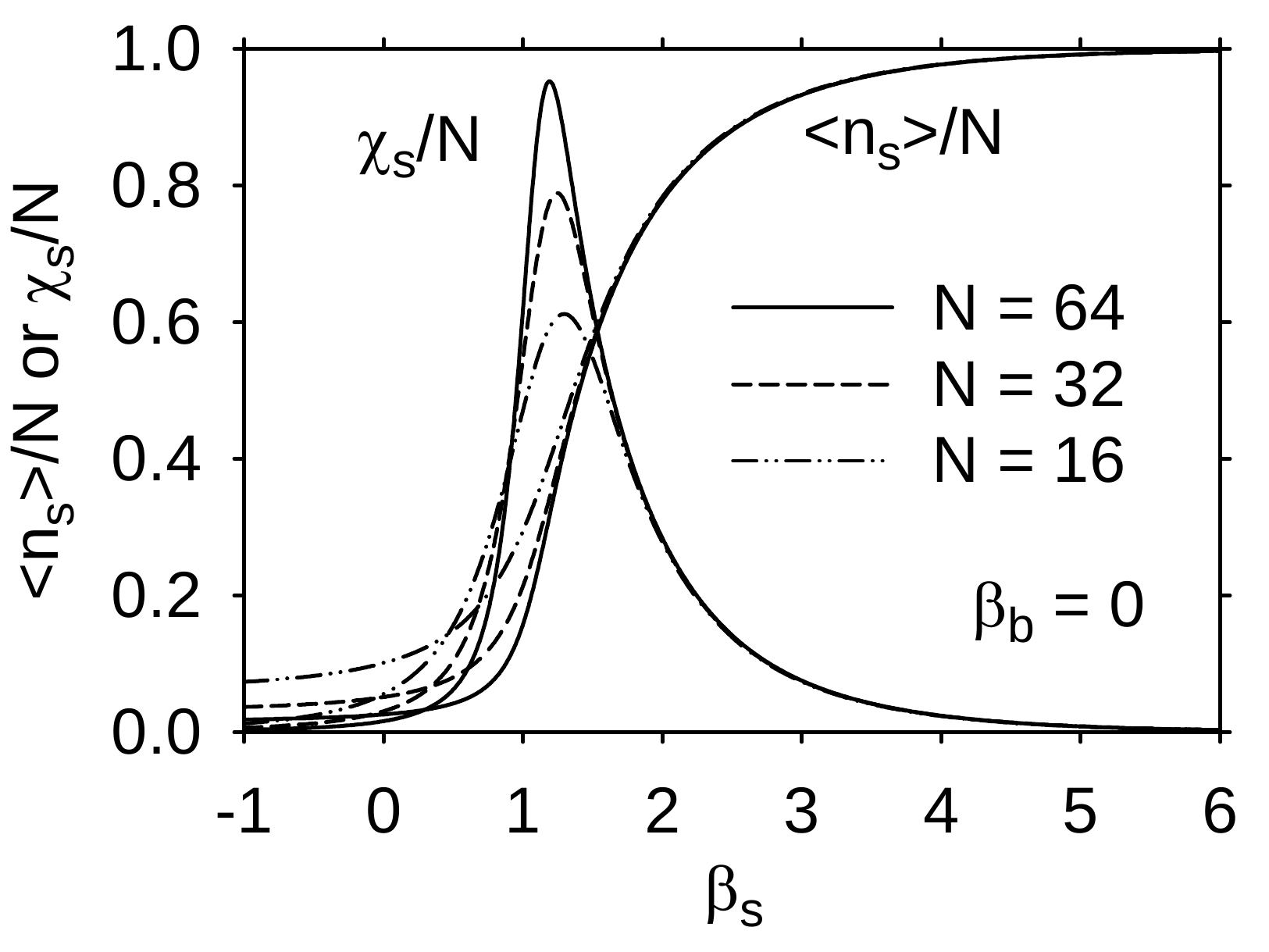}
\caption{\label{ns_and_chis}
Average number of surface contacts per monomer $\langle n_s\rangle/N$
and surface contact fluctuations, $\chi_s/N$, as a function of the
surface field $\beta_s$ for good solvent conditions ($\beta_b=0$). 
The lines represent results from the evaluation of the density of
states for  chains of length $N=64$ (solid), 32 (dashed), and 16
(dash-dotted), respectively. 
The graphs for $\langle n_s\rangle/N$ are monotonously increasing,
those for $\chi_s/N$ have a maximum in the transition region.}
\end{figure}

To explore adsorption in good solvent conditions, the bead contact
field is set to $\beta_b=0$ so that chain
segments interact with each other only through excluded volume
interactions. Density of states and production results are
evaluated as described in Sec.~\ref{eval_dos} for a range of surface
fields $\beta_s$.  
In Fig.~\ref{ns_and_chis} we present results 
for the average number of
surface contacts and their fluctuations as a function of the surface
field $\beta_s$.  
For negative $\beta_s$ values, the surface is repulsive and the number
of surface contacts is near $1$, the smallest possible value. As
$\beta_s$ increases, the surface becomes increasingly more attractive
and the number of surface contacts increases until it approaches $N$,
the largest possible value. 
The transition between desorbed and adsorbed states becomes steeper
with increasing chain length which is an expected finite size effect. 
The fluctuations $\chi_s$ are small for repulsive and very
attractive surfaces and have a maximum in the transition region.  
The location of the maxima are $\beta_s = 1.29$, 1.24, and 1.19 for
chains of length $N=16$, 32, and 64, respectively. 
We use maxima in $\chi_s$ to identify the transition fields for
surface adsorption in Sec.~\ref{phase_diags}.

\begin{figure}[t]
\includegraphics*[width=3.2in]{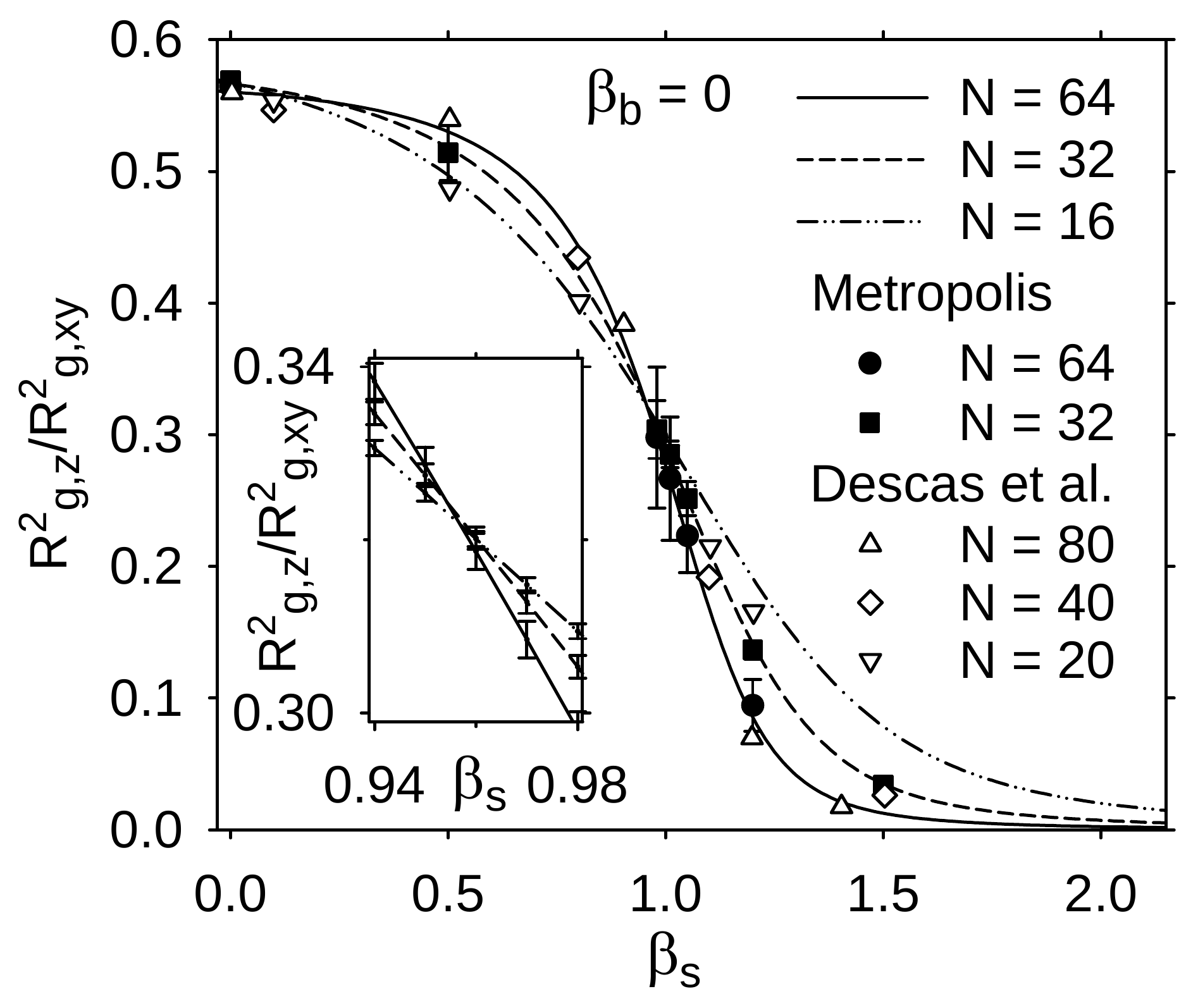}
\caption{\label{chaindimratio_bb0}
Ratio $R^2_{g,z}/R^2_{g,xy}$ 
of perpendicular and parallel contributions to the square radius
of gyration as a function of the surface field $\beta_s$ for good
solvent conditions ($\beta_b=0$).
The lines represent results from the evaluation of the density of
states and production data for chains of length $N=64$ (solid), 32
(dashed), and 16 (dash-dotted); the inset shows an enlargement of the
region where the lines cross. For clarity, error bars for our 
calculated values are shown only in the inset. (The
error bars generally increase with increasing $N$ and decreasing
$\beta_s$; for $\beta_s = 0$ they are about twice as large as for the
$\beta_s$-range of the inset).
The filled symbols with error bars represent Metropolis Monte Carlo
results from this work. The open symbols
represent Metropolis Monte Carlo results for chains of length $N=20$,
40, and 80 by Descas {\it et al.} \protect\cite{de04}. }
\end{figure}

The shape of a tethered chain may be described with the contributions 
$R^2_{g,z}\equiv R^2_{g,\perp}$ and $R^2_{g,xy}\equiv
2R^2_{g,\parallel}$ to the radius of gyration $R^2_g$ \cite{ei82}. 
For repulsive or
weakly adsorbing surfaces, the chain assumes mushroom configurations
whose extensions  parallel and perpendicular to the wall are
comparable, $R_{g,\perp}\gtrsim R_{g,\parallel}$. 
As the surface becomes more attractive, 
$R_{g,\perp}$ decreases and $R_{g,\parallel}$ increases, so that the
ratio $R^2_{g,z}/R^2_{g,xy} = 0.5R^2_{g,\perp}/R^2_{g,\parallel}$
decreases rapidly in the transition region. 
A scaling analysis shows that the ratio is independent of chain length
at the adsorption transition \cite{ei82}. 
In Fig.~\ref{chaindimratio_bb0} we present results for the ratio 
$R^2_{g,z}/R^2_{g,xy}$ as a function of $\beta_s$ for $\beta_b=0$. 
The lines represent results from an evaluation of our density of
states and production results for chains of length $N=16$, 32, and
64. 
For comparison, we have also included Metropolis Monte Carlo
results obtained by us for chains of length $N=32$ and 64, and by
Descas {\it et al.}\cite{de04} for chains of length $N=20$, 40, and
80. The agreement between data obtained with different methods is
good. 
While longer chains are required for a detailed comparison with
scaling predictions, our results for 
$R^2_{g,z}/R^2_{g,xy}$ show reasonable behavior. 
The lines for 
$N=64$, 32, and 16 intersect each other at about $\beta_s = 0.96$, as
shown in the inset. Our value for the intercept, 
$\beta_s = 0.96 \pm 0.02$, is consistent with the results of Descas 
{\it et al.}\cite{de04}, 
who determined an intercept of $\beta_s = 0.98 \pm 0.03$ from 
data which included chains up to length 200.  

\subsection{Chain collapse for a hard surface}\label{collapse}

A polymer chain undergoes a transition 
from extended to compact conformations, the coil-globule transition, 
when the solvent quality changes from good to poor. 
For the bond-fluctuation model 
with attractive bead-bead interactions, $\epsilon_b < 0$, 
this transition may be induced by reducing the temperature. 
The continuous coil-globule transition is
followed by a discontinuous crystallization transition at lower
temperature. 
For the model we are studying, which has a range of
attractive interaction $\leq \sqrt{6}$, the difference between the transition
temperatures decreases with chain length and, in the limit
$N\rightarrow\infty$, the transitions merge into a single 
first-order transition \cite{ra05b}.

\begin{figure}[tb]
\includegraphics*[width=3.2in]{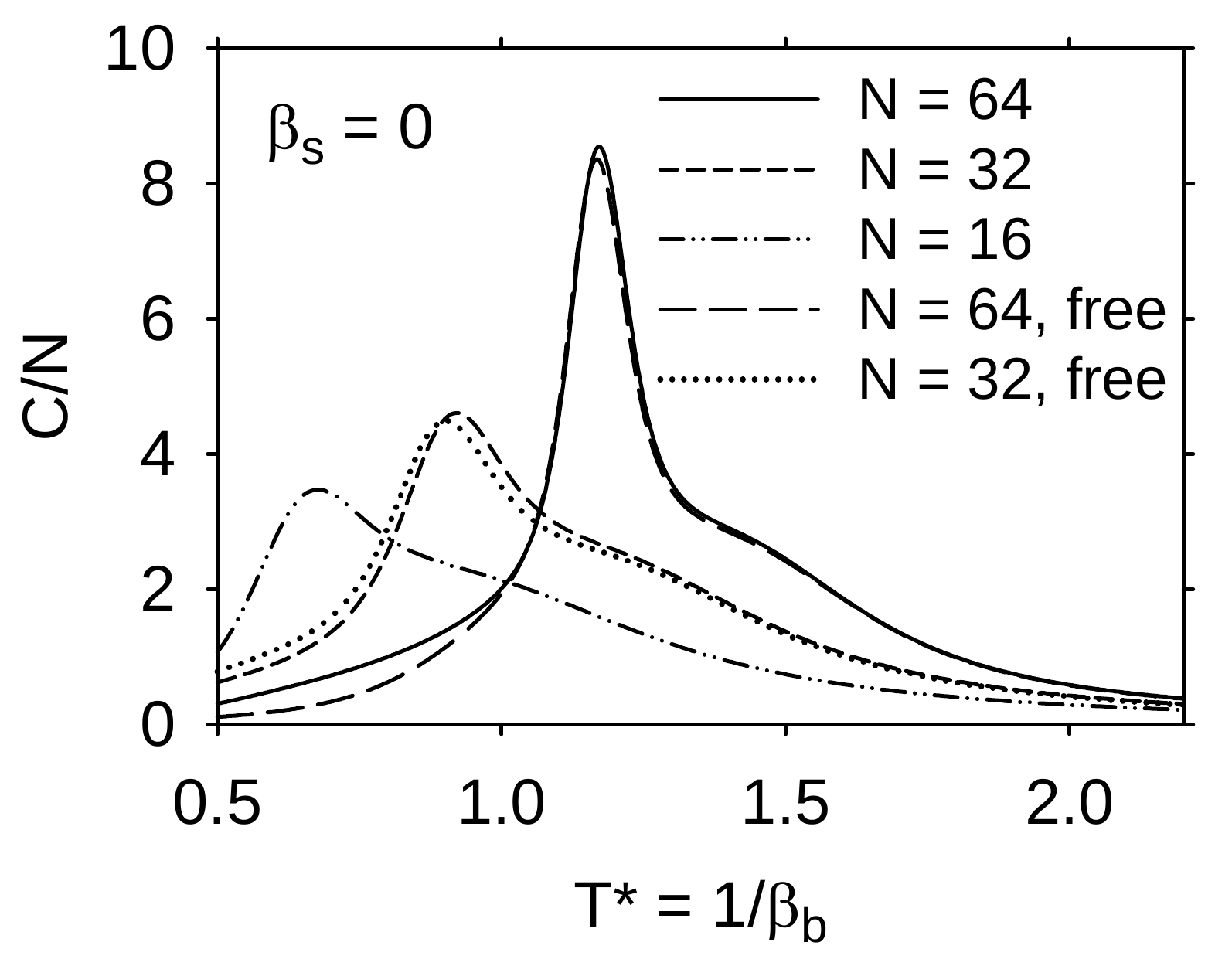}
\caption{\label{heat_caps}
Heat capacities per monomer, $C/N$, as a function of the reduced
temperature $T^* = 1/\beta_b$. 
The solid ($N=64$), short-dashed ($N=32$), and dash-dotted 
($N=16$) lines represent results for chains tethered to a hard surface
($\beta_s = 0$). 
The long-dashed ($N=64$) and dotted ($N=32$) lines represent results
for free chains \protect\cite{ra06b,ra05b}.
The estimated uncertainties of the results for the tethered chains are
smaller than the line thickness, except for $N=64$ at very low
temperatures, $T^* < 0.7$, where they correspond to about twice the
line thickness.}
\end{figure}

To explore chain collapse for tethered chains, we consider first 
a hard surface and set $\beta_s=0$. 
In Fig.~\ref{heat_caps} we present results for the heat capacity, $C$,  
as a function of the reduced temperature 
$T^* = 1.0/\beta_b$ obtained by evaluating our density of states
results for chains of length $N=16$, 32, and 64. 
For comparison, we also present results for the heat capacity of free
chains of length $N=32$ and $N=64$, obtained from Wang-Landau
simulations of the BF model \cite{ra06b,ra05b}. 
For each chain length considered here, the heat capacity shows a peak
at low temperatures followed by a shoulder at higher temperatures. 
For free chains, the peak is associated with the crystallization 
transition and the shoulder marks the coil-globule transition;   
we find the same to be true for the tethered chains 
In Fig.~\ref{config2d}~(a) we present an example for a compact and highly
ordered conformation of a chain of length $N=64$ that is typical for
the lowest temperatures.

While the heat capacity values for the tethered and free chains coincide 
at high temperatures, there are some differences at lower temperatures. 
For chains of length $N=64$ and temperatures smaller than 
$T^*\approx 1$, the free chain heat capacity values are lower than
those of the tethered chain.
This is due to a difference in the simulations; for the
free chain, a cutoff of 272 bead contacts was used while the tethered
chain results include bead contacts up to $n_{b,max}=288$. 
For somewhat higher temperatures, $1 \lesssim T^* \lesssim 1.5$, 
there is also a small temperature shift (barely visible in the graph)
between the heat capacity curves of the free and tethered $N=64$ chains. 
For chains of length $N=32$ and temperatures smaller than about 
$T^*= 1.2$, i.e.\ below the coil globule transition, the 
heat capacity curve for the free chain is shifted
to lower temperatures compared to that of the tethered chain. 
A comparison of the normalized density of states results for the free
and tethered chains of length $N=32$ 
shows that the number of available conformations
decreases with increasing bead contact number more rapidly for free
chains than for tethered chains. This suggests that 
the hard surface reduces the relative number of extended conformations
more than the relative number of compact conformations. Since
the fraction of monomers belonging to the surface of a conformation
decreases with chain length, it seems reasonable that the effect
is larger for shorter chains; in our simulations, the effect has all
but disappeared for $N=64$. 
In the limit $N\to\infty$, we expect 
the transition temperature for the three-dimensional
coil-globule transition to be the same for free and tethered chains. 

\begin{figure}[tb]
\includegraphics*[width=3.2in]{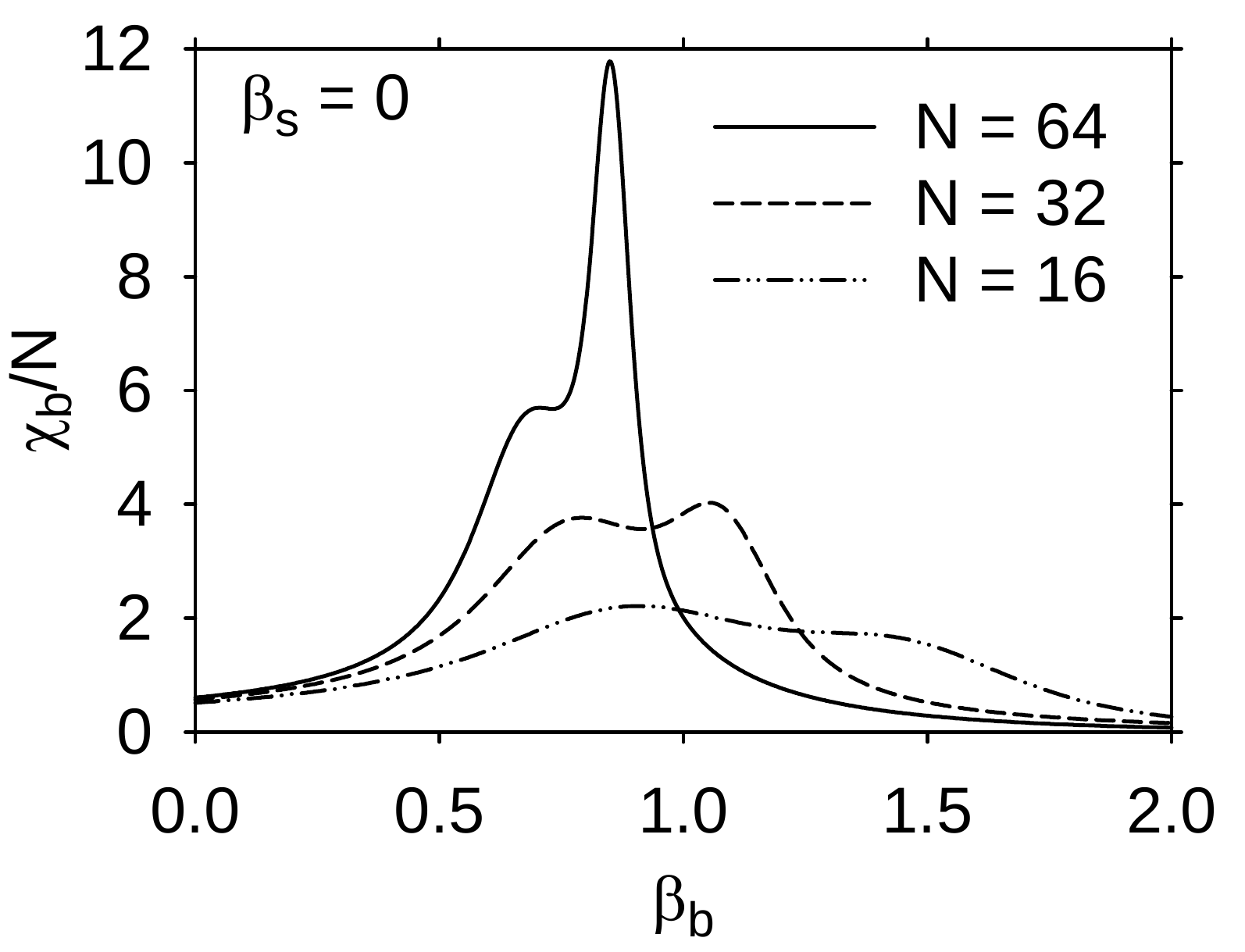}
\caption{\label{chib_bs0}
Monomer-monomer contact fluctuations, $\chi_b/N$, as a function of the
field $\beta_b$ for a hard surface, $\beta_s = 0$. 
The lines represent results from the evaluation of the density of
states for tethered chains of length $N=64$ (solid), 32 (dashed), and
16 (dash-dotted). }
\end{figure}

To determine the location of the transitions it is 
convenient to consider the fluctuations in the number of bead
contacts $\chi_b$. These are related to the heat capacity through
Eq.~(\ref{heatcap}). 
In Fig.~\ref{chib_bs0} we present 
tethered chain results for $\chi_b$ as 
function of the field $\beta_b$ for chains of length $N=16$, 32, and
64 at $\beta_s=0$.
The graphs for $N=32$ and $N=64$ show two well separated maxima 
while the graph for $N=16$ has one maximum associated with the
coil-globule transition and a shoulder associated with the 
globule-globule transition. 
The locations of the maxima for the coil-globule transitions are 
0.91, 0.79, and 0.71 for chains of length $N=16$, 32, and 64,
respectively. 
The field values for the globule-globule transitions are
1.47 (estimated), 1.05, and 0.850 for $N=16$, 32, and 64
respectively. 
The values for the globule-globule transitions compare well with those 
obtained from the peaks in the heat capacity, which yield 
$\beta_b$ values of 1.48, 1.08, and 0.853 for $N=16$, 32, and 64,
respectively. 
We use maxima in $\chi_b$ to identify the transition fields for
chain collapse in Sec.~\ref{phase_diags}.

The transitions of single untethered polymer chains have been 
investigated intensively with a variety of off-lattice polymer models 
\cite{mi93,zh96d,zh97,li00e,ta01b,ca02d,ma03,pa06b,pa06c,se08}.  
Except for the shortest chains and the most short-ranged interactions,
the heat capacity as a function of 
temperature shows two prominent features; a peak at higher
temperatures that indicates the transition from disordered coil
conformations to disordered globule conformations, and a peak at low
temperatures that indicates the transition from a liquid-like
disordered globule to a solid-like ordered globule. 
Typically, the peak associated with the order-disorder transition,
which is sometimes referred to as the ``freezing'' transition, 
sharpens with increasing chain length and is accompanied by a bimodal
probability distribution as expected for a discontinuous transition. 
Our results for  tethered and free chains show that the chain collapse
in the bond fluctuation model is analogous to chain collapse in
off-lattice models. 

\subsection{Chain collapse in two dimensions}\label{collapse_2d}

\begin{figure}[tb]
\includegraphics*[width=3.2in]{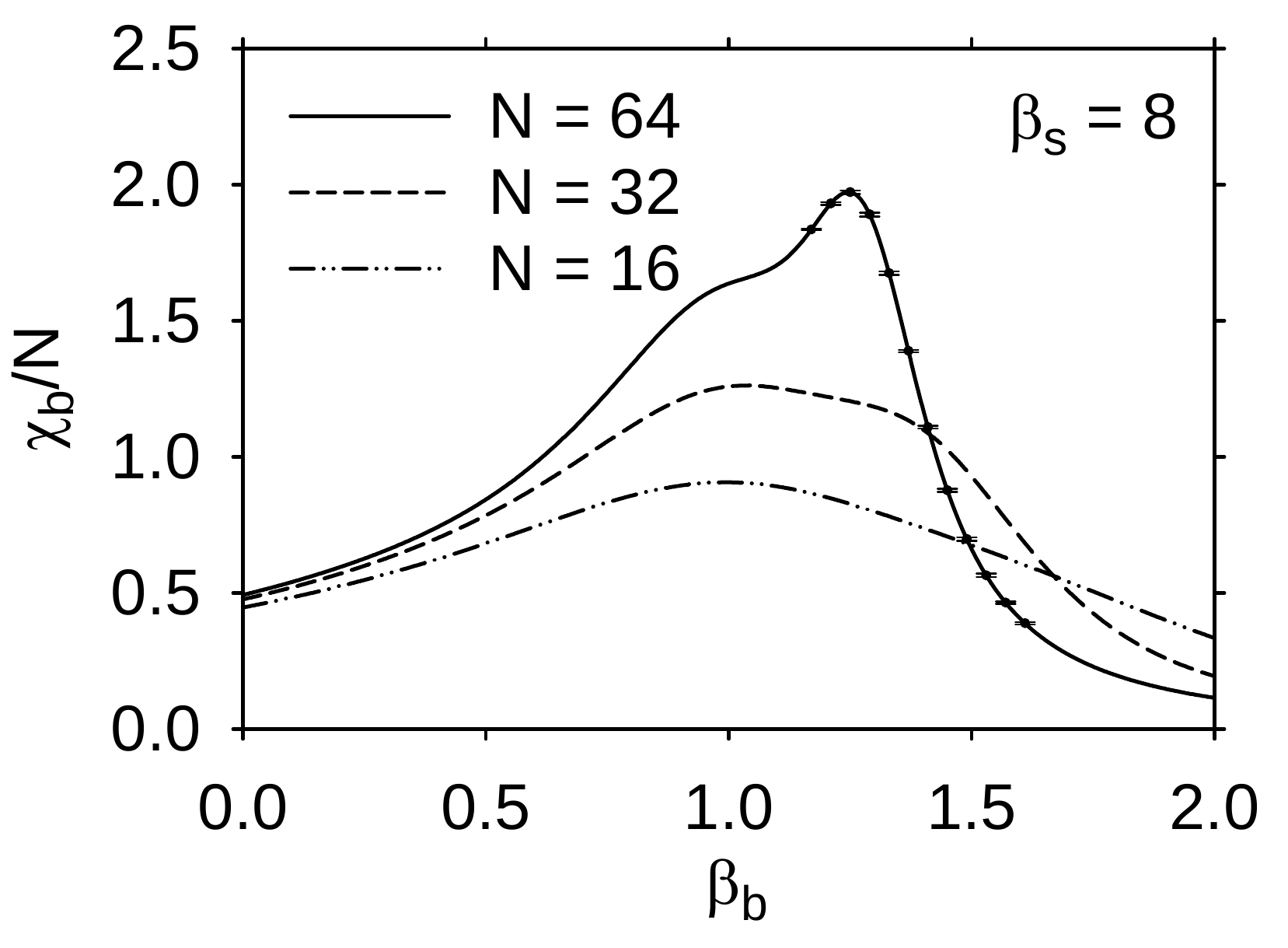}
\caption{\label{chib_bs8}
Monomer-monomer contact fluctuations, $\chi_b/N$, as a function of the
field $\beta_b$ for a very attractive surface, $\beta_s = 8$. 
The lines represent results from the evaluation of the density of
states for tethered chains of length $N=64$ (solid), 32 (dashed), and
16 (dash-dotted). }
\end{figure}

\begin{figure}[tb]
\includegraphics*[width=3.2in]{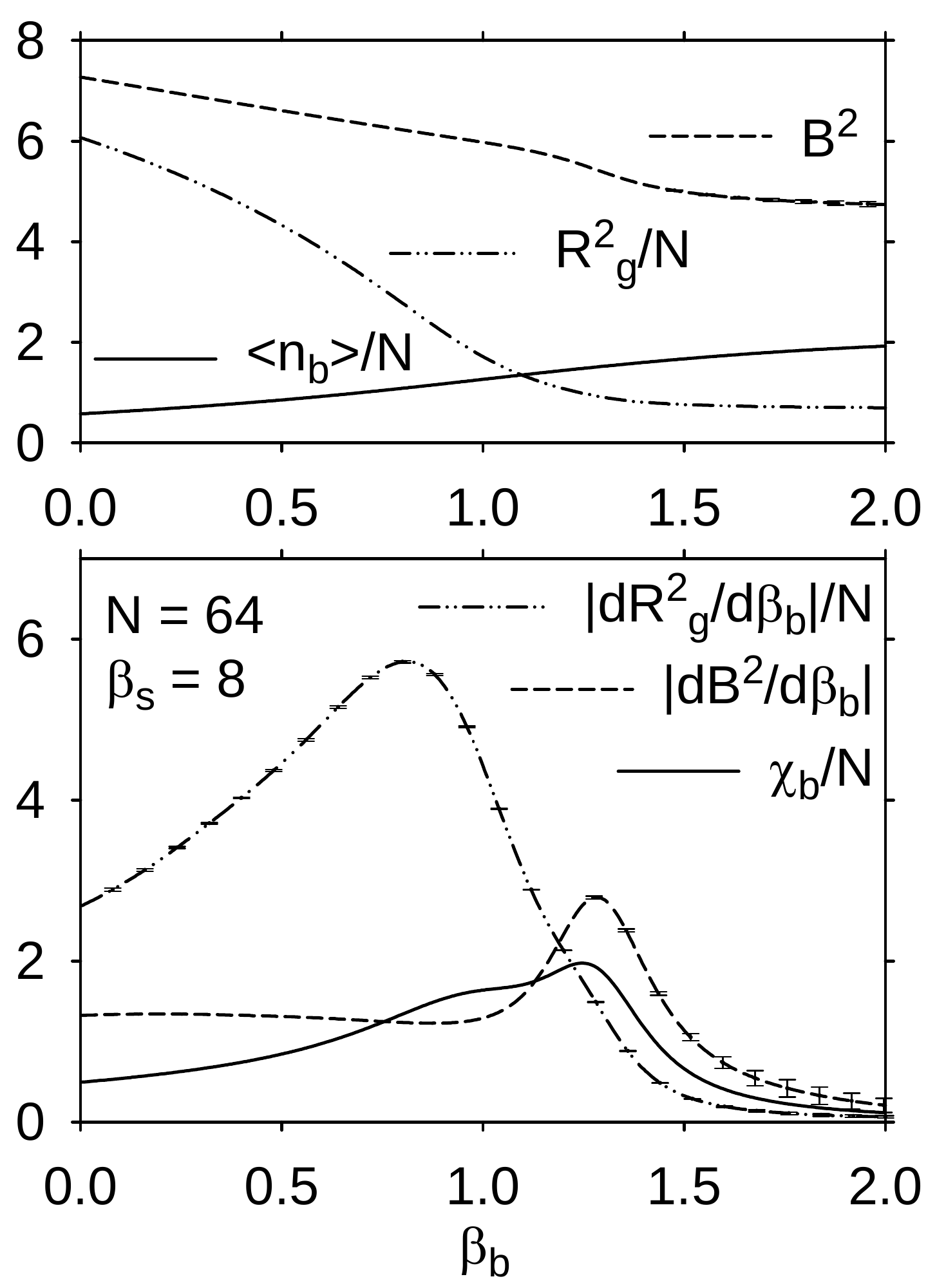}
\caption{\label{dims_and_derivs_bs8}
Chain collapse for large surface field. The top panel shows the 
average square bond lengths $B^2$, radius of gyration $R^2_g$ divided
by the chain length $N$, and
the average number of monomer-monomer contacts per bead $\langle
n_b\rangle/N$ as a function of 
the monomer-contact field $\beta_b$ for a chain of length
$N=64$ and a surface field $\beta_s = 8.0$.
The bottom panel shows how these quantities change as the chain
undergoes the collapse transition. 
The dashed and dash-dotted lines represent absolute values of the
numerical derivatives $(\partial R^2_g/\partial \beta_b)_{\beta_s}$ 
and $(\partial B^2/\partial \beta_b)_{\beta_s}$, respectively.
The solid line represents $\chi_b/N$ as in
Fig.~\protect\ref{chib_bs8}. }
\end{figure}

If a chain is completely adsorbed to a surface, a collapse transition
between two-dimensional states may be observed.
To investigate this transition, we chose a strong surface field, 
$\beta_s = 8$, and evaluated the density of states as usual, with 
probabilities given by Eq.~(\ref{prob}). 
For comparison, we also performed calculations 
using only density of states results with the maximum number of
surface contacts, $n_s=N$, and evaluated them with 
probabilities given by Eq.~(\ref{proba}). 
For the range of $\beta_b$ values relevant to the collapse transition,
the results are indistinguishable. (For much higher $\beta_b$ values,
the transition to layered states appears in the evaluation of the full
density of states but not the restricted set.) 
In Fig.~\ref{chib_bs8} we present results for the bead contact
fluctuations $\chi_b$ of tethered chains as a function of $\beta_b$  
for chains of length $N=16$, 32, and 64 at $\beta_s = 8$. 
For $N=16$, only a single maximum is discernible, while for $N=32$ the
maximum is broad and accompanied by a shoulder on the high $\beta_b$
side. For $N=64$, a narrower peak in $\chi_b$ is preceded by a
shoulder on the low $\beta_b$ side. In analogy with the two-stage
collapse transition for chains tethered to a hard surface, we expect
the features at low and high $\beta_b$ values to indicate 
coil-globule and globule-globule transitions in two dimensions,
respectively.

\begin{figure}[t]
\includegraphics*[width=2.0in]{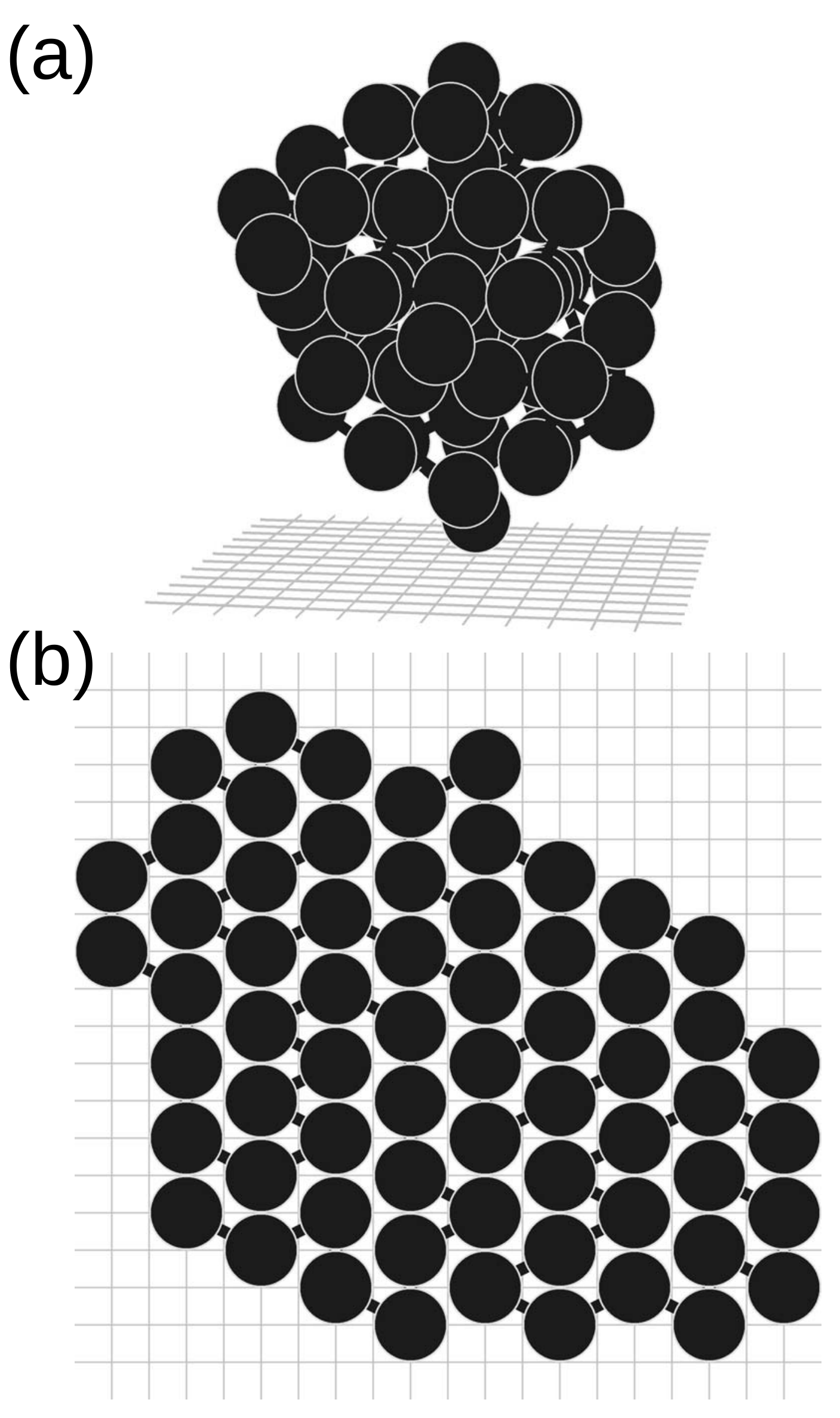}
\caption{\label{config2d}
Examples for compact conformations of chains of length $N=64$. 
(a) A highly ordered three-dimensional conformation representative of
the desorbed compact (DC) region of the phase portrait in
Fig.~\protect\ref{phase_diagN64} 
(b) A highly ordered two-dimensional (single layer) conformation 
representative of the adsorbed compact (AC) region of the phase
portrait. 
In both diagrams, the size of the circles corresponds to the hard core
diameter of the beads; the bonds are shown as wide lines. }
\end{figure}

To investigate this further, we calculated the square radius of
gyration $R_g^2$ and bond lengths $B^2$ of the chains. 
In Fig.~\ref{dims_and_derivs_bs8} we present results for 
these quantities and their derivatives with respect to $\beta_b$ as a
function of $\beta_b$ for a chain of length $N=64$ in the surface
field $\beta_s=8$.   
For comparison, the average number of bead contacts, $\langle
n_b\rangle/N$, and their fluctuations, $\chi_b/N$, are also shown. 
The graphs show that the largest changes in the radius of gyration
occur  near the shoulder of the $\chi_b$ graph, while 
the largest changes in the bond length occur
near the maximum of $\chi_b$. 
Since the radius of gyration measures the overall size of a chain
conformation, while the bond lengths represent a small-scale
property, the results suggest that a coil-globule transition 
is followed by a local rearrangement of segments. 
As in the three-dimensional case, the states beyond this latter
transition are highly ordered, i.e., we interpret this
``globule-globule transition'' as a transition from a fluid-like to a
crystal-like state. In Fig.~\ref{config2d}~(b)
we show an example for such a highly ordered two-dimensional 
conformation of a chain of length $N=64$. 
As expected for an order-disorder transition, we find a bimodal
probability distribution for the occupation of states at the
globule-globule transition. 

A two-stage transition from a disordered extended coil through a
disordered globule to an ordered globule in two dimensions has also
been observed in simulations with a parallel tempering
algorithm of an off-lattice (bead-spring) model \cite{ma03}. 
The symmetry of the two-dimensional ordered conformations is
model dependent. 
For the BF model employed in this work, 
the unit cell for the ordered interior of Fig.~\ref{config2d}~(b) has
basis vectors ${\bf a}_1 = 2a\hat{y}$ and  
${\bf a}_2 = a\hat{y} + 2a\hat{x}$, where $\hat{x}$ and $\hat{y}$ are
the unit vectors in the $x$ and $y$ direction, respectively, and $a$
is the lattice constant. 
The number of nearest neighbors (six) is the same as for a hexagonal
lattice, which is the symmetry for the two-dimensional ordered state
of the off-lattice model \cite{ma03}. 
For the ISAW model, on the other hand, 
compact two-dimensional chain conformations have the symmetry of the
square lattice with four nearest neighbors. 
It is an open question to what extent these kinds of two-dimensional
states carry over to chemically realistic models including rough
substrates.  

\subsection{Phase portraits}\label{phase_diags}

\begin{figure}[t]
\includegraphics*[width=3.2in]{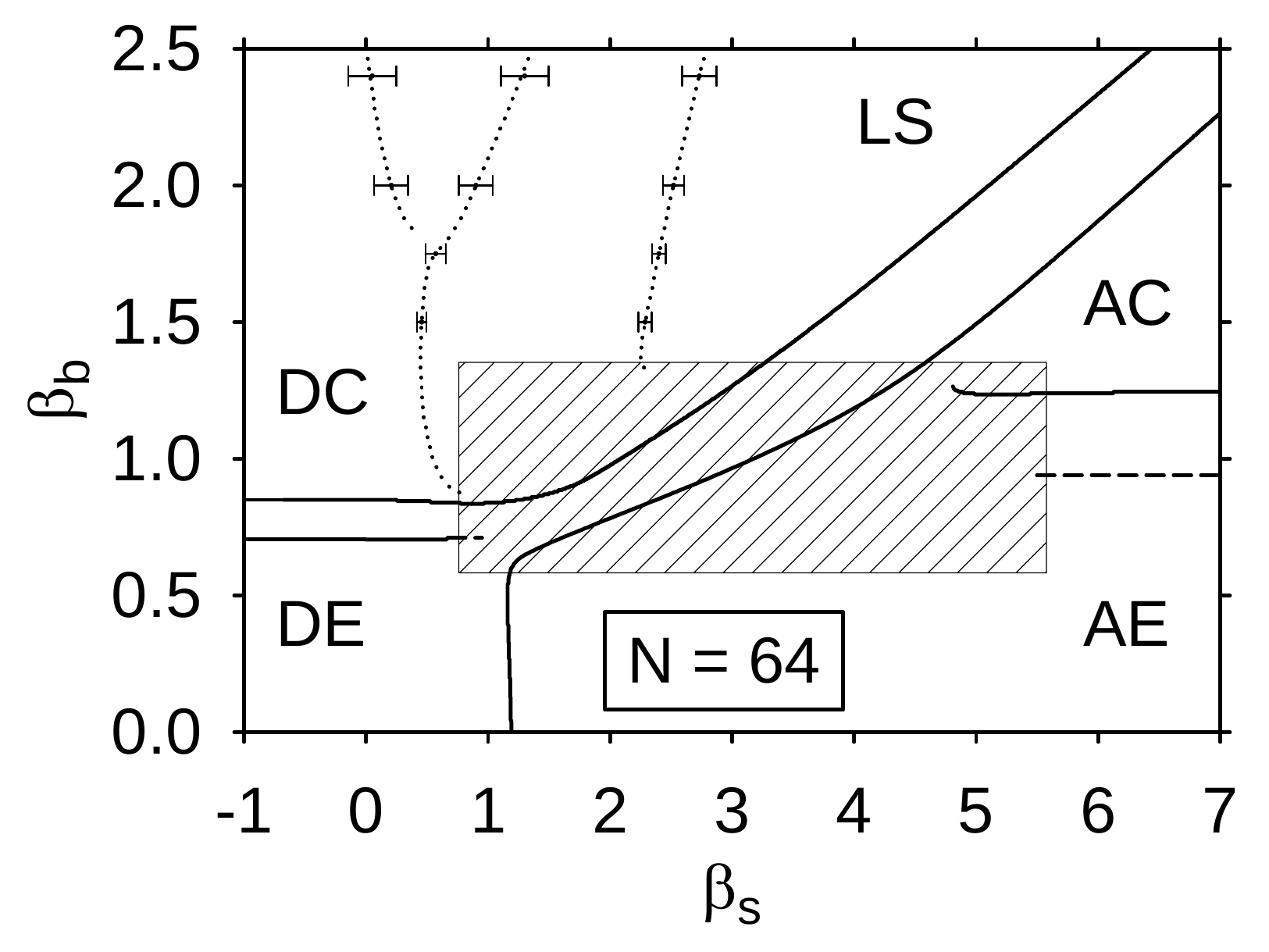}
\caption{\label{phase_diagN64}
Phase portrait for a tethered chain of length $N=64$
 in the space of the surface field ($\beta_s$) and
bead-contact field ($\beta_b$). The regions are named as in
 Fig.~\protect\ref{fig1}.  
The solid lines represent maxima of 
surface ($\chi_s$) and bead-contact ($\chi_b$) fluctuations, as
 explained in the text.  
The dashed lines are an estimate for the location of the coil-globule
 transition from the ``shoulder'' on the susceptibility $\chi_b$.  
The dotted lines represent shallow maxima in the susceptibility
 $\chi_s$ that depend sensitively on the details of the available
 compact chain conformations.
In the shaded area near the center of the diagram, the susceptibility
``landscapes'' are too complex to identify all of the maxima
clearly. This is why some of the lines end rather than merge with
other lines. }
\end{figure}

In the preceding sections we have employed maxima in the
susceptibilities $\chi_s$ and $\chi_b$ to identify 
transition fields $\beta_s$ and $\beta_b$ 
for chain adsorption in good solvent and for chain
collapse near hard and very attractive surfaces. 
In this section, we extend this approach to a wide range of
conditions and construct a phase portrait for chains of length
$N=64$, shown in Fig.~\ref{phase_diagN64}, that may be compared with
the phase diagram of Fig.~\ref{fig1} discussed in the Introduction. 

From the maxima and ``shoulders'' of the susceptibilities we
identified transition lines in the space of field variables $\beta_s$
and $\beta_b$ that separate the desorbed extended (DE),
adsorbed extended (AE), desorbed compact (DC), adsorbed compact
(AC), and layered state (LS) regions. 
The shaded area near the center of the phase portrait in
Fig.~\ref{phase_diagN64} marks a range of
field values where the susceptibility
``landscapes'' are too complex to identify all of the maxima
clearly. This is the reason why some of the lines end rather than
merge with other lines. In this discussion, we focus on 
field values outside the shaded area. 

In the diagram of Fig.~\ref{phase_diagN64}, 
the horizontal solid lines separating the DE and DC regions represent
maxima of $\chi_b$ associated with the dual collapse transition in
three dimensions.  
Similarly, the horizontal solid and dashed lines 
separating the AE and AC regions represent maxima and ``shoulders'' of
$\chi_b$ associated with the collapse transition in
two dimensions. 
For both collapse
transitions, the fluctuations in 
the number of surface contacts, $\chi_s$, are not significant. 
The vertical solid line separating the DE and AE regions represents
maxima in $\chi_s$ associated with adsorption in good solvent. For
this transition, the fluctuations in the number of bead contacts,
$\chi_b$, are small. 
The two solid lines that separate the regions of adsorbed states (AE 
and AC) from fully or partially desorbed states in poor solvent
conditions (DC and LS) 
represent maxima of both surface and bead-contact fluctuations. 
The dotted lines in the LS region represent shallow maxima of
$\chi_s$. The error bars show the spread of the location of 
these maxima when different sets of density of states values are
evaluated. The transitions
associated with these maxima are structural, discontinuous transitions
that depend sensitively on the details of the available 
compact chain conformations. 

The AC states for high $\beta_s$ and $\beta_b$ values have
strictly two-dimensional, single-layer conformations with almost 
hexagonal symmetry as shown in Fig.~\ref{config2d}~(b). The chain
conformations for slightly lower $\beta_s$ values, in the region
between the two solid lines near the upper right corner of
Fig.~\ref{phase_diagN64}, have exactly two layers, each with nearly
hexagonal symmetry and shifted against each other by one lattice
constant $a$ of the underlying cubic lattice.
The transition from single to double layers is similar in nature to
the layering transition observed in the ISAW model. 
In the limit of infinite chain length, 
the slope of the line separating the two-layer from the one-layer
conformations may be determined by equating the energy
per segment of the single and double layer conformations
\cite{kr05}. For the ISAW model, the value of this slope is 1/2. 
For the bond-fluctuation model employed in this work, 
the calculated value for infinitely long chains is 1/3 while finite
size effects increase it to a value
of 0.38 for the chains of length $N=64$. 
 States in the region marked LS in
Fig.~\ref{phase_diagN64} are very different from the single and
double-layer states discussed above. The chain conformations are
three-dimensional with a symmetry group corresponding to a cubic
lattice with basis. The number of surface contacts decreases stepwise
through the transitions marked by dotted lines until only a single
surface contact is left in the region marked DC. 

For the model of a tethered chain investigated in this work, all
transitions from good to poor solvent conditions are dual in
nature. 
The character of the transitions, continuous or discontinuous, and
their dependence on the surface field $\beta_s$ 
will be discussed in more detail in a later publication.
Of course, the transitions between different states of a finite chain
are no sharp phase transitions, as they occur for systems in the
thermodynamic limit, but rather gradual changes in the weights of various
microstates of the model system.
With increasing chain length, the separation
between the transition lines decreases so that a phase portrait
resembles more closely the phase diagram of Fig.~\ref{fig1} as the
chain length increases. 
In agreement with theoretical predictions for the ISAW~\cite{vr96}, we
find that the  
transition line describing the coil-globule transition in good solvent
is horizontal and perpendicular to the adsorption transition
line. 
In contrast to the ISAW model, where the crystalline phase always has 
simple cubic symmetry and the most compact conformation is cubic, the
BF model supports more than one ordered phase and the most compact
conformation is highly faceted. This leads to a complex 
sequence of transitions in the poor solvent regime, which we will
investigate in more detail in later work. 

\subsection{Relation to real polymers near surfaces}\label{relation}

Transitions in polymers near surfaces may be induced by changing the
solvent quality or, more typically, by changing the temperature. 
Since the BF model employed in this work orders without forming 
lamellae and since polymer crystallization is typically dominated by
kinetic effects \cite{ch05} a direct comparison with available
experimental data on surface crystallization of polymers is not
possible. However, we will discuss some qualitative aspects of ordered
chains near surfaces.  
Experiments on alkanes physisorbed on graphene surfaces at low
coverage show the chains to be rod-like (all trans conformations) and
oriented parallel to the surface \cite{he92,ar02}. 
Experiments on thin films of a variety of crystallizable polymers show
that the orientation of the chains relative to surface depends on the
thickness of the layer (see, for example,
Ref.~[\onlinecite{wa04b}]). For very thin films, one typically finds
``edge-on'' lamellae corresponding to chains with their
backbones oriented perpendicular to the surface. For example, for 
poly(ethylene oxide) (PEO) films on bare silicon wafers, crystallites 
with chain backbones perpendicular to the surface grow from a layer of
adsorbed, non-crystallized chains oriented parallel to the surface
through a partial dewetting of the surface \cite{re00b,so00}. 

The fields $\beta_b$ and $\beta_s$ employed in this work combine
the inverse temperature $\beta$ and the interaction parameters for net 
monomer-monomer interactions $\epsilon_{b}$ and monomer surface
interactions $\epsilon_{b}$, respectively (see Eq.~(\ref{fields})). 
In our phase portrait of Fig.~\ref{phase_diagN64}, 
the origin corresponds to the limit of infinite temperature while
a range of temperatures at constant interaction parameters 
corresponds to a segment of a straight line through the origin with
the slope given by the ratio 
$\epsilon_b/\epsilon_s$ of the interaction parameters. 
For example, for systems without attractive surface interactions (hard
surface) a decrease in temperature moves the system up along the 
vertical line with $\beta_s = 0$ in Fig.~\ref{phase_diagN64} and leads
to the two-stage transition  from desorbed extended to desorbed
compact states.  
When the monomer-monomer interaction parameter $\epsilon_b$ has a much
smaller magnitude than the surface interaction parameter $\epsilon_s$,
a decrease in temperature moves the system along an almost horizontal
line from $(0,0)$ to higher $\beta_s$ values in
Fig.~\ref{phase_diagN64} and leads to the adsorption transition from
desorbed (DE) to adsorbed (AE) extended chain conformations. 
For systems where the interaction parameters $\epsilon_b$ and
$\epsilon_s$ are comparable in size, our phase portrait predicts
transitions from the disordered DE states to the ordered
three-dimensional LS states upon lowering the temperature. 
While the details of the LS states clearly depend on the model,
layered states are not uncommon. Off-lattice simulations
of confined polymers, for example, show layered states for strong
confinement \cite{ma03} and the adsorption of small molecules from the
vapor onto solid substrates leads to the occurrence of layered
structures (``multilayer adsorption'').
When the surface interactions are somewhat larger than the
monomer-monomer interactions, a decrease in temperature is
expected to induce chain adsorption to the disordered AE states,
followed by transitions to the ordered LS states.
During the ordering, the chain partially desorbs from the
surface to reach the LS states. 
A combination of partial dewetting and ordering in a
three-dimensional crystallite is also observed in
experiments on PEO on silicon \cite{re00b}. Finally, 
when the surface interactions are very strong, a transition between
two-dimensional disordered (AE) and ordered (AC) states is expected from
our calculations. For real polymeric systems, such strong surface
attractions might be provided by specific interactions (for example
hydrogen bonding) between the surface and the polymer.

While some polymers undergo a crystallization transition as the
temperature is lowered there are many polymers that remain
amorphous. For example, in atactic poly(butadiene) the random 
distribution of monomers of different stereoregularity prevents the
crystallization of the chains. 
In order to describe non-crystallizable polymers, one introduces 
an element of randomness in the composition of the model polymer. In this
case, we expect the phase portrait in field space to be qualitatively
the same as shown in Fig.~\ref{phase_diagN64} for low $\beta_b$
values, but to contain single 
lines of continuous transitions between disordered
extended and compact states. In addition, there may be minor
transitions in the compact regions corresponding to local reordering
of sections of a chain.

\section{Summary and Conclusions}\label{summary}

In this work we performed 
Monte Carlo simulations of a bond-fluctuation model for a tethered
chain with two-dimensional Wang-Landau algorithms and umbrella sampling. 
The simulations yield density of states results that have been
evaluated for interaction parameters spanning the range
from good to poor solvent
conditions and from repulsive to strongly attractive surfaces. 
For given fields $\beta_s$ and $\beta_b$, we calculated expectation
values for chain dimensions, contact numbers for surface and
monomer-monomer contacts, and fluctuations in the contact numbers. 
Three types of transitions were investigated in some detail. 
For adsorption in good solvent, we compared our results with
Metropolis Monte Carlo data for the same model and found good
agreement. 
For the collapse transition in three dimensions, we considered 
chains tethered to a hard surface and found them to behave very
similar to free chains, 
with the differences between the two
situations decreasing with increasing chain length, as expected.
For the collapse transition in two dimensions, we found a dual
transition with the same sequence of transitions 
that is observed for three-dimensional chains: 
a coil-globule transition that changes the
overall chain size is followed by a local rearrangement of chain
segments.  

In order to investigate the overall phase behavior of 
the tethered chains considered in this work, 
we located maxima of the susceptibilities
$\chi_s$ and $\chi_b$ in the $\beta_s$--$\beta_b$ plane. 
We found that all transitions from good to poor solvent conditions are
dual in nature and that the separation between the two lines 
belonging to the same transition decreases with increasing chain
length.  
In agreement with theoretical predictions for the ISAW
model\cite{vr96}, we find that the  
transition line describing the coil-globule transition in good solvent
is horizontal and perpendicular to the adsorption transition
line. 
For poor solvent conditions, early work on the ISAW model 
predicted the existence of a surface attached globule (SAG) phase
\cite{si01,ra02b,mi03} which was later found to consist of a whole
sequence of layered states \cite{kr05}. In the corresponding parameter
region we find one transition (between single and double layer states)
that is similar to the AC to LS (SAG) transition of the ISAW. The next
transition of the BF model, however, changes the symmetry of the
ordered segments and has no correspondence in the ISAW model. 
From the maxima of the susceptibilities of finite-length chains 
alone it is not possible to resolve the nature of the transitions near
points where transition lines meet. 
This is an interesting question that we will try to
address in future work.

\section*{Acknowledgments}

Financial support through the Deutsche Forschungsgemeinschaft (grant
No. SFB 625/A3) and a sabbatical leave from the
University of Akron are gratefully acknowledged. 

\appendix

\section{Simulation details}\label{details}

In this section, we present details of the density-of-states and
production simulations for chains of length $N=16$, $N=32$, and
$N=64$. 
For each chain length, density of states results were
generated with Wang-Landau algorithms and then 
refined with umbrella sampling. In order to compare results from
groups of simulations, the average over the $\ln(g)$ values in 
the interior ($N/4 < n_s < 3N/4$, $5N/16 < n_b < 25N/16$) of the range
of accessible states is calculated for each simulation and the
$\ln(g)$ values are shifted by this amount. To determine
uncertainties for the WL results, the average and standard deviation
of the $\ln(g)$ values are calculated for each state $(n_s,n_b)$. 
For umbrella sampling results, the log-density of states values are
weighted by the length of the umbrella sampling simulation and
the weighted average $\overline{\ln(g)}$ and standard deviation
$\sigma_{\ln(g)}$ are calculated. 
Parameters for the umbrella sampling simulations and some numerical
characteristics of the final densities of states are presented in
Table~\ref{table_dos}. In the following, we highlight
differences in the simulations for the different chain lengths. 
Production was carried out either in separate simulations or
concurrently with the construction of the density of states. A summary
of the production simulation parameters is presented in
Table~\ref{table_production}. 

\begin{table}[ht]
\caption{Umbrella sampling parameters and some density of states
  characteristics for chains of length $N=16$, 
  $N=32$, and $N=64$. $N_u$ is the number of results for which
  umbrella sampling simulations were performed, $K_u$ is the length
  of the simulations in Monte Carlo steps, 
  $N_s$ is the number of states sampled, 
  $n_{b,\mathrm{max}}(1)$ and $n_{b,\mathrm{max}}(N)$
  represent the largest number of bead-bead contacts for $n_s=1$ and
  $n_s=N$ surface contacts, respectively. The values for the range of
  the log-density of states ($\Delta \ln(g)$), and its uncertainties
  ($\sigma_{\ln(g)}$) represent results after umbrella sampling.  For
  $N=16$ and 
  $N=32$, the states in this table are believed to be the complete
  set; for $N=64$, only states included in our evaluation
  are represented. 
\label{table_dos}}
\vspace*{0.5cm}
\begin{ruledtabular}
\begin{tabular}{|l|r|r|r|}
$N$ & 16 & 32 & 64 \\ \hline 
$N_u$ \rule{0em}{1.2em} & 4 & 3 & 2 \\
$K_u/10^9$ & 1 & 
16, 10, 18 & 18.5, 18.8 \\
$N_s$ & 748 & 3467 & 15268 \\
$n_{b,\mathrm{max}}(1)$ & 53 & 128 & 288 \\
$n_{b,\mathrm{max}}(N)$ & 34 & 76 & 163 \\
$\Delta \ln(g)$ & 50.8 & 109.5 & 227 \\
average $\sigma_{\ln(g)}$ & 0.008 & 0.008 & 0.015 \\
median $\sigma_{\ln(g)}$ & 0.004 & 0.003 & 0.004 \\
maximum $\sigma_{\ln(g)}$ & 0.18 & 0.5 & 1.5000 \\
\end{tabular}
\end{ruledtabular}
\end{table}

\begin{table}[ht]
\caption{Production parameters for chains of length $N=16$, $N=32$,
  and $N=64$. The table entries represent the simulation length in
  MC steps, where configurations are evaluated every 10 MC steps. 
  The left column indicates the type of simulation. For simulations
  sampling with the density of states, the number of 
  replicas ($N_r$) is indicated and, in one case, the maximum number 
  of bead contacts considered. For simulations with the Metropolis
  acceptance criterion, the values for the field variables $\beta_s$
  and $\beta_b$ are shown. 
\label{table_production}}
\vspace*{0.5cm}
\begin{ruledtabular}
\begin{tabular}{|l|r|r|r|}
$N$ & 16 & 32 & 64 \\ \hline 
$N_r=1$ \rule{0em}{1.2em} 
& $1\times 10^8$ & $4\times 10^9$ & $3.35\times 10^9$ \\
$N_r=3$ & $1\times 10^9$ & &  \\
$N_r=4$, $n_b\leq 256$  &  &  & $2.6\times 10^8$ \\
$N_r=5$ &  & $1\times 10^9$ & $3\times 10^8$ \\
$\beta_s = \beta_b = 0$ & $1\times 10^9$ & $1\times 10^9$ & $1.5\times 10^9$ \\
$\beta_s = \beta_b = -1$ &  &  & $1\times 10^9$ \\
$\beta_s = \beta_b = -2$ & $2\times 10^9$ & $1\times 10^9$ & $4\times 10^8$ 
\end{tabular}
\end{ruledtabular}
\end{table}

$\bm{N=16}$
For chains of length $N=16$, four simulations with the original
Wang-Landau algorithm were performed. The histogram was
considered flat when the visits to each individual $(n_s,n_b)$ state 
were no less than 0.8$\overline{h}$, where $\overline{h}$ is the
average of the visits to all states.
The simulations required between 1.2$\times 10^9$ and  1.9$\times
10^9$ Monte Carlo steps to converge. 
The results agreed well with each other; the average, median, and maximum 
standard deviation of the $\ln(g)$ values are 0.017, 0.008, and
0.63, respectively. The largest deviations occur 
for states with the largest
number of bead contacts for given number of surface contacts, i.e.\
for states near the upper rim of the range of accessible states 
in Fig.~\ref{dos_range}. 
Umbrella sampling simulations with $10^9$ Monte Carlo steps were
performed for each of the WL density of states results for $N=16$ and
evaluated as described in Sec.~\ref{umbrella}.
The agreement between the four different results is
excellent, as may be seen from the values of the standard deviations
presented in Table~\ref{table_dos}. As before, the largest deviations
occur at the upper rim of the range of accessible states.

$\bm{N=32}$
For chains of length $N=32$, simulations with the original Wang-Landau
algorithm would not converge. We modified the flatness criterion and
considered averages over groups of states. Each group consisted of $N$
states and two sets of groups were created. One set was formed by
going up along columns in the $(n_s,n_b)$ plane and adding states to a
group until it was filled, continuing in the next column when the top
was reached before the group was complete. The other set was formed by
going along rows in the same way. The histogram was considered flat 
when the visits to each group of states in the two sets was no less
than 0.8 and no more than 1.25 of the average of the visits over all
groups. 
We obtained results from two simulations (A and B) 
with a Wang-Landau algorithm where the density of states was updated
only for accepted moves (see Eq.~(\ref{update})) and 
from one simulation (C) with a five-replica Wang-Landau algorithm
where the density of states was updated after accepted as well as
after rejected moves, Eqs.~(\ref{update}) and (\ref{updateb}).
The results for the density of states from the five replicas agreed
well in the regions of overlap and were combined with a simple linear
switching function. 

The $N=32$ simulations took between $1.3\times 10^{10}$ and $4.1\times
10^{10}$  MC steps to converge. All simulations missed or severely
underestimated the density of states values of four states on the
upper rim of accessible states in Fig.~\ref{dos_range}. 
Except for states on the rim, the two single-replica results A and B 
agree reasonably well with each other, 
the average, median, and maximum standard deviation of the $\ln(g)$
values are 0.029, 0.011, and 4.4, respectively. 
However, there are systematic deviations between the $\ln(g)$ values
for high $n_b$-states between result C from the
five-replica simulation and results A and B.  

In order to prepare for umbrella sampling, the density of state
results for $N=32$ from the Wang-Landau 
simulations were augmented by assigning reasonable guesses to the
$\ln(g)$ values that were missed or nearly missed in the
simulations. 
We performed umbrella sampling simulations with the parameters given
in Table~\ref{table_dos} and updated the density of states results as
described in Sec.~\ref{umbrella}. 
The systematic deviations between results A, B and C disappeared after
umbrella sampling and the density of states values agree very well,
(see the values for the standard deviations in Table~\ref{table_dos})
except for states very close to the upper rim of the range of
accessible states in Fig.~\ref{dos_range}. 
For results A and B, umbrella sampling led to significant increases
in the normalized $\ln(g)$ values for a large range of states with
high numbers of bead contacts. 
For result C, umbrella sampling decreased the normalized $\ln(g)$ values 
slightly for states with very high numbers of bead contacts and
changed the values significantly on the rim of accessible states. 
One state was not visited in the C simulation and its density of state
value was calculated from the A and B results only.
Production simulations were carried out with the parameters in
Table~\ref{table_production}. 
For some states near the rim, production data were supplemented by
evaluating stored chain conformations.

$\bm{N=64}$
For chains of length $N=64$, one density of states result with a
cutoff in the number of bead contacts at $n_b=4N$ was obtained
with the modified Wang-Landau algorithm described for chains of length
$N=32$ (simulations A and B), above. This simulation was completed in
$2\times 10^9$ MC steps; simulations without cutoff would not converge. 

In order to obtain density of states results for a larger range of
bead contact values, we applied the global update algorithm described
in Section \ref{global}.
For the results presented here, we used the following shift parameters
in Eq.~(\ref{shift}), 
$\kappa=\lambda=\omega=10^4\ln(f_k)$, 
where $\ln(f_k) = 2^{-(k-1)}$ is the increment of the log density of
states at refinement level $k$. 
The first global update simulation 
used a value of $\rho = 10^2\ln(f_k)$ for the uniform growth
criterion. 
It was started at the $k=15$ level and allowed to  
progress through about $8\times 10^8$ MC steps before the
refinement level was set to $k=18$. At this level, about $7.5\times
10^8$ MC steps were carried out before our groups-of-state flatness
criterion was satisfied and the simulation progressed through the final two
levels in a standard Wang-Landau algorithm with our flatness
criterion, which took about $3.5\times 10^8$ MC steps, for a total
simulation time of about $1.9\times 10^9$ MC steps.
The density of states values obtained in this global update simulation
agreed reasonably well in the range of common states with the results 
obtained in the simulation with cutoff described above. It extended
the range of visited states to higher bead contacts but it missed a
number of states with large surface and bead contact values that had
been found in the simulation with cutoff. After removing states 
whose density of states values had been severely underestimated,
we combined the results from
the two simulations with a linear switching function; this combined
density of states is referred to as result A in the following.

The second global update simulation was started at the $k=18$
level and used a value of $\rho = 10\ln(f_k)$. With these parameters,
the density of states in the interior is generated through a larger 
number of smaller shifts. The simulation was allowed to proceed
through a total of about $3.15\times 10^{10}$ MC steps, after which time
we collected the density of states, our result B. 
During this time, global updates and the addition of new states became
increasingly rare. For example, only 31 of the more than 15000 visited
states were found in the second half of the simulation. The range of
visited states from this simulation is larger than that of result A.
The differences between results A and B are largest 
for states with very high bead-contact
values; the average, median, and maximum standard deviation of the
$\ln(g)$ values are 0.16, 0.1, and 18.3.
In order to prepare for umbrella sampling, the density of state
results A and B were smoothed by assigning reasonable guesses to the
$\ln(g)$ values that were missed or nearly missed in the simulations. 

Unfortunately, the umbrella sampling simulations for both A and B
missed some states completely (20 for A and 12 for B out of a
total of 15275 input states), nearly missed a few others (8 for A and 2
for B), and found some states for the very first time (12 for A and 14
for B, with 4 in common for a total of 22 new states). 
All of the problematic states are very close to the upper rim of the 
range of accessible states shown in Fig.~\ref{dos_range}.
Except for the newly found states, which were discarded, the $\ln(g)$
values were updated at the end of the umbrella sampling as described
in Section~\ref{umbrella}. 
For states that were found in both umbrella sampling simulations, 
the log-density of states results were combined as described above. 
For states that had been missed or nearly missed in one of the
umbrella sampling simulations, we used the density of states result
from the other simulation and assigned an uncertainty of
$\sigma_{\ln(g)}=1.5$.
Except near the rim, the agreement between the results is very good
(see Table~\ref{table_dos}). 
Production simulations were carried out with the parameters in
Table~\ref{table_production}. 
For some states near the rim, production data were supplemented by
evaluating stored chain conformations.  For 13 states on the rim,
values of the quantities of 
interest were estimated by extrapolation from data with smaller
numbers of bead contacts and generous errors were assigned.
The impact of the rim states depends on the values of the field
variables, in particular on the value of $\beta_b$, as discussed in
Section~\ref{dos_results}. In this work, we focus 
on conditions where their impact is small. For future evaluation, 
we are currently performing simulations with a slightly modified
simulation algorithm that we hope will improve the density of states
results near the rim.

\end{document}